\documentstyle[aps,eqsecnum,epsf,preprint,tighten]{revtex}
\newcommand{\gras}[1]{\mbox{\boldmath $#1$}}
\newcommand{\be}{\begin{equation}}
\newcommand{\ee}{\end{equation}}
\newcommand{\ba}{\begin{eqnarray}}
\newcommand{\ea}{\end{eqnarray}}
\newcommand{\Mc}{{\cal M}}
\newcommand{\Ms}{M_{\odot}}
\newcommand{\m}{\langle}
\newcommand{\M}{\rangle}
\newcommand{\hf}{\tilde h}
\newcommand{\rf}{\tilde r}

\newcommand{\bml}{\begin{mathletters}}
\newcommand{\eml}{\end{mathletters}}
\def\ltsima{$\; \buildrel < \over \sim \;$}
\def\simlt{\lower.5ex\hbox{\ltsima}}
\def\gtsima{$\; \buildrel > \over \sim \;$}
\def\simgt{\lower.5ex\hbox{\gtsima}}
\newcommand{\COCCIAGWE}{{\em Gravitational Wave Experiments},
edited by E. Coccia, G. Pizzella and F. Ronga
(World Scientific, Singapore, 1995)}
\begin{document}
\draft

\title{ Gravitational waves from coalescing binaries and Doppler experiments}
\author{Bruno Bertotti}
\address{
Dipartimento di Fisica Nucleare e Teorica, Universit\`a di Pavia\\
27100 Pavia, Italy}
\author{Alberto Vecchio}
\address{Max Planck Institut f\"{u}r Gravitationsphysik, 
Albert Einstein Institut\\
14473 Potsdam, Germany}
\author{Luciano Iess}
\address{Dipartimento di Ingegneria Aerospaziale, Universit\`a
"La Sapienza"\\
00158 Roma, Italy}

\date{\today}

\maketitle

\begin{abstract}

Doppler tracking of interplanetary spacecraft provides the only method
presently available for broad-band searches of low frequency gravitational 
waves 
($\sim 10^{-5}-1\,{\rm Hz}$). The instruments have a peak sensitivity around 
the reciprocal of the round-trip light-time $T$ ($ \sim 10^3 - 10^4$ sec) of the 
radio link connecting the Earth to the space-probe 
and therefore are particularly suitable to search for coalescing binaries 
containing massive black holes in galactic nuclei. 
A number of Doppler experiments -- 
the most recent involving the probes ULYSSES, GALILEO and MARS OBSERVER -- 
have been carried out so far; moreover,  in 2002-2004 the CASSINI spacecraft 
will perform three 40 days data  acquisition runs with expected sensitivity 
about twenty times better than that achieved so far.

Central aims of this paper are: (i) to explore, as a function of the relevant 
instrumental and astrophysical parameters, the Doppler output produced by {\it in-spiral} signals -- 
sinusoids of increasing frequency and amplitude (the so-called {\it chirp}); (ii) 
to identify the most 
important parameter regions
where to concentrate intense and dedicated data analysis; (iii) to
analyze the all-sky  and all-frequency sensitivity of the CASSINI's experiments,
with particular emphasis on possible astrophysical targets, such as our Galactic 
Centre and the Virgo Cluster. 

We consider first an ideal situation in which the spectrum of the noise is white and there are 
no cutoffs in the instrumental band; we can define an 
{\it ideal } signal-to-noise ratio (SNR) which depends in a simple way on the fundamental 
parameters of the source -- {\it chirp mass} 
$\Mc$ and luminosity distance -- and the experiment -- round-trip light-time 
and noise spectral level. 
For any {\it real} experiment we define the {\it sensitivity function} $\Upsilon$ as the degradation 
of the SNR with respect to its ideal value due to a
coloured spectrum, the experiment finite duration $T_1$, the accessible frequency band 
($f_b, f_e$) of the signal and the source's location in the sky. 
We show that the actual value of $\Upsilon$ crucially depends on the overlap of the  
band $(f_b,f_e)$ with the instrument response: the sensitivity is best when $f_b \simlt 1/T$ 
and $f_e$ coincides with the frequency corresponding to the beginning of the merging phase. 
Furthermore, for any $f_b$ and $T_1$, there is an
optimal value of the chirp mass -- the {\it critical chirp mass} 
$\Mc_c \propto f_b^{-8/5}\,T_1^{-3/5}$ -- that produces the largest sensitivity function; lower 
values of 
$\Mc$ correspond to a smaller bandwidth and less SNR. Also the optimal source's location
in the sky strongly depends on $(f_b,f_e)$.

We show that the largest distance at which a source is detectable with CASSINI's experiments is $\sim 600$ Mpc 
and is attained for massive black holes of comparable masses $\sim 10^7\,\Ms$ and 
$f_b \sim 10^{-5}\, {\rm Hz}$. Sources not far from coalescence  in the Virgo cluster with 
$10^6 \, \Ms\simlt \Mc\simlt 10^9\, \Ms$ would be detectable with a SNR $\sim 1-30$. The SNR and 
the range
of accessible masses reduce drastically when a smaller mass ratio 
is considered. We then turn the attention to galactic observations,
in particular on the detectability of a coalescing binary in the Galactic Centre, where a small 
black hole of mass $M_2$ could be orbiting around the central massive one 
$M_1 \simeq 2\times 10^6\,\Ms$. CASSINI would be able to pick up such systems 
with $M_2 \simgt 50\,\Ms$; for $M_2 \simgt 10^3\,\Ms$ the SNR could be as high as $\sim 100
- 1000$. It may also be possible to detect such binaries in more than one of the three CASSINI 
experiments, thus re-enforcing the confidence of detection. 

\end{abstract}

\bigskip

\newpage
\narrowtext \onecolumn

\section{INTRODUCTION}
\label{sec;intr}

The search for gravitational waves has greatly grown in the
last few years: four ground based laser interferometers --
LIGO \cite{LIGO}, VIRGO \cite{Virgo}, GEO600 \cite{GEO} and TAMA \cite{TAMA} -- 
with optimal sensitivity
in the band $\sim 10\,{\rm Hz} - 1\,{\rm kHz}$ are now under
construction and are scheduled to be in operation by the turn of the century;
at the same time, the sensitivity
of acoustic narrow-band devices in the ${\rm kHz}$ regime is
steadily improving \cite{Nautilus,Explorer,Auriga,LSUbar,Blairbar}. In this high frequency band 
the most interesting sources are connected to collapsed objects
with mass in the range $\sim 1 - 1000\, \Ms$. However, 
at low frequencies (below a few {\rm Hz}) detectors on Earth
are severely impaired by the seismic noise. To
observe sources of higher mass one must use instruments in space, sensitive in the ${\rm mHz}$ band.

The Doppler tracking of interplanetary spacecraft \cite{EW} is the only method presently available
to search for gravitational waves 
in the low frequency regime ($\sim 10^{-5} - 1\,{\rm Hz}$). Several 
experiments have been carried out so far, all of them with {\it non dedicated} space-probes:
 VOYAGER 1 \cite{voy}, PIONEER 10 and 11 \cite{pio10,pio11}, 
ULYSSES \cite{pap0,pap1}, 
GALILEO and MARS-OBSERVER \cite{UMOGcoinc}. They suffer
from the limitations of instrumentation designed for other
purposes and from several disturbances, including those due to the ground
sector; however, their sensitivity is astrophysically interesting, even 
though the rate of detectable events is uncertain and probably low, given the current
instrument performances. At present a
large amount of data is available: a major 
experiment lasted 28 days was carried out with ULYSSES in 
February/March 1992 \cite{pap1} and
a 20 days coincidence experiment was performed in the Spring
1993 with the spacecraft  GALILEO, MARS-OBSERVER and
ULYSSES \cite{UMOGcoinc} (see Table \ref{tab;doppexp}). 

The space-probe CASSINI -- a NASA/ESA/ASI joint mission \cite{cassini} -- 
represents the next step in Doppler experiments: launched on October 15th 1997 
with primary target the study of the Saturn system, the spacecraft carries on board
much improved instrumentation and 
will perform three long (40 days each) data acquisition runs in 2002, 2003 and 2004 
to search for gravitational waves \cite{cas} with expected sensitivity
about twenty times better than that achieved so far.
The next far away and much more ambitious projects will probably involve
interferometers in space with arms of millions of km \cite{LISA,OMEGA}
(see also \cite{BBgr14} and references
therein for a review on detectors in space).

In the low frequency band we expect five main types of sources (for an
extensive discussion regarding the whole
gravitational wave spectrum we refer the reader to 
\cite{Thorne95,Thorne97}):

\begin{enumerate}

\item Catastrophic, wide-band collapses of large masses, possibly in
dense concentrations of matter and stars, leading  to the formation of a massive black hole
(MBH);  the search for such signals motivated the original proposal 
of Doppler experiments \cite{TB}.

\item Short-period binary systems of solar mass compact objects
\cite{LP,LPP,HBW}; among them the binary pulsar PSR 1913+16 \cite{Hulse94,Taylor94}, whose study
over the past 25 years has provided the most clear {\it indirect evidence} of the existence of
gravitational waves \cite{TW89}; however, stellar mass binaries 
are not detectable with present and near-future Doppler experiments, due 
to inadequate sensitivity and short duration.

\item Solar mass compact objects orbiting a massive black
hole in galactic cores \cite{HB,SR,Shibata}.

\item Binary systems of massive  black holes spiraling together
toward their final coalescence \cite{BBR,Hae,Lisavecch}.

\item Backgrounds of gravitational waves of primordial origin
or generated by the superposition of radiation coming from
unresolved binaries \cite{CA,GV,BH,PP98,KP98}.

\end{enumerate}

Here we will consider signals emitted by binary systems containing
massive black holes (MBH's), therefore sources at point 3 and 4, and the main goals of the paper
are the exploration of the structure of the Doppler output produced by in-spiral signals from coalescing
binaries and the analysis of the all-sky/all-frequency sensitivity of Doppler experiments
with emphasis on the CASSINI mission.

Compelling arguments suggest the presence of MBH's
in the nuclei of most galaxies, as the near-inevitable byproduct 
of the infall of gas in their potential well \cite{Rees93,HR93,LR94} and
MBH's are invoked to explain a number of phenomena, such as the activity 
of quasars and active galactic nuclei \cite{ZN64,Salpeter64,Blandford90,Osterbrock93}. 
However, the observational evidences of their existence come mainly from observations
of relatively nearby galaxies, whose nuclei do not show significant activity. 
Central dark masses -- that are believed to be "dead
quasars" \cite{Rees90} -- are mainly inferred by studying the spatial distribution and velocity of gas and/or 
stars in galactic cores. About ten candidate massive black holes, with mass in
the range $\sim 10^6 - 10^9\,\Ms$, are known today and we refer the reader to
\cite{KR95,KR96,Marel96,Rees97} for recent reviews.
The most striking evidences come from the mapping of the gas
motion via the 1.35 cm water maser emission line in NCG 4258
and the observations in the near infrared band 
of the star motion in Our Galactic Centre. In the case of NCG 4258, observations with the 
Very Long Baseline
Array have allowed to measure the velocity of the gas in the disk with an accuracy of 1 km/sec: the
disc rotates with a velocity exactly in agreement with Kepler low, due to the
presence of an obscure object of mass $\simeq 3.6\times 10^6\,\Ms$ \cite{Miyoshi95}. 
Regarding Our Galactic Centre,
recent observations with the New Technology Telescope of the proper
motion of stars in the core have shown that their speed scales as the inverse of the
square root of the distance from the center up to 20000 km/sec; the inferred mass of the
dark body is $\simeq 2.5 \times 10^6\,\Ms$ \cite{EG96}. In both cases
only exotic (and highly implausible) alternatives to a MBH can still be considered 
\cite{maoz96,maoz98}.

Many galaxies have experienced at least a merger since the epoch $z > 2$ \cite{BH92,Fritze}.
If a MBH is present in each core of both interacting
galaxies, the two massive objects would fall into the common potential well,
forming a pair which losses energy and angular momentum through dynamical friction; eventually a binary
forms and driven by radiation reaction precedes toward the final coalescence. 
This scenario provides, therefore, a quite natural way of producing
massive binary black holes (MBHB's) \cite{BBR}. The apparent
precession and bending of jets in active galaxies \cite{BBR,Roos} and 
Doppler shifted broad emission line peaks in quasars \cite{GaskellR} are possible
indirect evidences of the existence of such systems. Moreover observations of MBHB's
in the nuclei 1928+738 \cite{RKH}, 
OJ287 \cite{SHVSB,VVB} and 3C 390.3 \cite{Gaskell} have been claimed.

MBH's in the center of galaxies are surrounded by a dense cluster of 
ordinary stars, whose tidal disruption provides
the gas to refuel the central object,  white dwarfs, neutron stars and low mass back holes,
probably in the mass range $\sim 5 - 100\,\Ms$. Unless ordinary stars, compact objects
are not disrupted by the hole and can occasionally be captured, presumably on highly
eccentric orbits, forming binary systems with life shorter than the Hubble time. Moreover
the orbits of objects with mass $\simgt 5 \Ms$ around the central massive body would not
be perturbed by other stars in the core providing therefore clean gravitational wave
signals, that can be used to extract valuable information about the central object 
\cite{Poisson96,Ryan97}.

Taking the data analysis point of view, where the structure of the signal
determines the choice of the optimal algorithm, it is convenient
to divide the entire process of binary {\it coalescence}
in three separate phases:

(a) {\it Adiabatic in-spiral}: the bodies are driven by radiation
reaction from their initial distance to smaller separations, 
losing energy and angular momentum. The emitted
gravitational wave corresponds to a sinusoid of increasing frequency
and amplitude (the so-called {\it chirp}). In this
paper we will make two important assumptions: (i) when the signal
enters the sensitivity window of the detector, the orbit has already been
circularized; in fact, the
eccentricity $e$ diminishes with the frequency $f$ according to 
$e^2(t)\propto f^{-19/9}(t)$ \cite{Pet}. A binary born with a period of 100
years, say, even with $e \sim 1$ would carry a residual 
eccentricity $\simlt 10^{-4}$ when the signal can be picked up by the instrument. 
For binaries composed of two MBH's, that 
have undergone a common evolution inside a galactic core,
this assumption is reasonable, even-though the full understanding of 
massive binary evolution before radiation reaction takes over still lacks
\cite{MV,PR,VCP}.
The condition $e \ll 1$ is likely to be violated for solar mass compact
objects orbiting a massive one, where the secondary body is 
captured in a highly elliptic orbit, which 
maintains a not negligible eccentricity throughout the in-spiral \cite{HB}. 
This would bring about a different scenario, with shorter life-time
and emission of  gravitational waves at multiples of the orbital
frequency \cite{Wahl}. (ii) In the modelling of the waveform we will use the lowest-order
Newtonian quadrupolar approximation, as we are mainly aiming at the estimation of the signal-to-noise
ratio (SNR) and not at the construction of signal templates for matching 
filters \cite{Schutz89,Schutz97}.

(b) {\it Merger}: when the body separation $r$ reaches a value of about
$6\,M$, where $M$ is the source's total mass,
the orbit becomes unstable and
the binary begins the final deadly phase of plunge-in.
The signal consists in a wide-band 
burst of duration $\sim M$, but the details of the emitted radiation
are in large part still unknown.

(c) {\it Ring-down}: the single black hole formed during the 
merging phase settles down oscillating
according to quasi-normal modes. The emitted waves consist of a superposition of
exponentially damped sinusoids.

In this paper we will concentrate on radiation emitted during the in-spiral phase.
Such signals can be usefully classified according to the best
algorithm to extract them in a given data set \cite{pap1}:
\begin{description}
\item{ I.\hspace{.33cm} {\it Periodic signals}: the rate of change of the frequency 
is too small to be detected with the instrumental frequency 
resolution. Here the ordinary techniques for the
search of periodic oscillations of unknown frequency apply.}
\item{ II.\hspace{.23cm} {\it Linearly "chirped" signals}: the frequency drift is 
larger than the frequency resolution of the detector
but the change in frequency due to its second time derivative during the
observation time
can be neglected, so that we can assume a linear increase of $f(t)$
with a constant rate $\beta = df/dt$. An efficient technique to 
search for these signals has been developed in \cite{Smith,AAL,AT} and 
already applied to several data sets \cite{pap1}; 
for every allowed $\beta$, the chirping signal is reduced to
a periodic one and standard methods can be applied. }
\end{description}
These two classes correspond to sources which are far from
coalescence -- in fact $df(t)/dt\propto f^{-8/3}(t)$, see Eq. (\ref{fdot}) --
and, for a given distance, weaker; therefore, the search depth
(the maximum distance at which a binary would be
detectable) is very limited: in \cite{pap1} we showed that,
with signals of this class, the 
Galactic Centre is marginally accessible with the best data
available today. 
\begin{description}
\item{ III. {\it General chirps}: the assumed signal has the correct
evolution in frequency, but the merger does not occur inside the
record.}
\item{ IV.\hspace{.01cm} {\it Chirps and bursts}: the record contains the 
final part of the chirping waveform, and the wide-band burst 
produced just before the engulfment of 
one black hole into the horizon of the other.}
\end{description}
These classes of signals can be distinguished in the record by the 
frequency and its change; for a given frequency, its drift increases with 
the class. Roughly speaking, this ordering corresponds to 
stronger sources and, therefore, to larger attainable distances.

Although in the final phase of the in-spiral the two bodies have high 
velocities ($v/c> 0.1$) and the description of their gravitational wave
emission requires a relativistic approach \cite{Thorne95}
we shall use the quadrupole Newtonian approximation for the estimation
of the signal-to-noise ratio (SNR). The use of this approximation has the 
advantage that the in-spiral signal depends on 
only {\it one physical parameter}, the {\em chirp mass} $\Mc$,
see Eq. (\ref{cm}). In principle it can be derived 
from the measurement of the frequency $f$ and its rate of 
change $df/dt$; then the amplitude $A\propto \Mc^{5/3}\,f^{2/3}/D$
provides the distance $D$ of the source. 
We can therefore determine, for any given experiment, the largest attainable 
distance; we call it, for a given class, the {\it ken} of the 
experiment. This is another example, similar to the one pointed out by 
Schutz in \cite{Schutz86}, of the power of gravitational 
wave astronomy when a simple source model is available. 
Central aim of the paper is to fully develop the richness of these concepts
when applied to Doppler experiments and 
to link possible astrophysical sources (like those in the Virgo cluster
and/or in the Galactic Centre) to the actual data. 
In particular we will apply the main results to the set of experiments
scheduled during the CASSINI's mission.    

Doppler tracking of interplanetary spacecraft provides a wide band detector of 
gravitational waves in the low frequency regime (but it can also be used as narrow-band
xylophone instrument \cite{Tinto}). A very stable electromagnetic signal in the GHz radio band
is continuously transmitted from the Earth to the spacecraft and coherently 
transponded back in order to monitor the change of their relative velocity. 
The optimal sensitivity of the detector is at frequencies of the order of
the inverse of the round-trip light-time $T$ of the Earth-spacecraft radio link.
However, since in practice the duration of the experiment is not large, only a limited 
frequency  interval of a chirping signal is accessible. 
There are two modes of search for a signal: if the frequency interval is small, {\it narrow band 
search}, one looks for sources far from the final plunge-in, hence at a small distance. 
In a {\it wide band search} the signal is stronger and one can detect radiation emitted
by binary systems further away.
In the past, the analysis of Doppler data has been confined mainly to the narrow band 
case (classes I and II); in this paper we concentrate on the more interesting  
wide band case and aim at sources at larger distances, but detectable only for a 
smaller fraction of their lifetime. 
The largest "ken" is, of course, attained for class IV, which merges into 
the class of generic, wide-band bursts.
The search for bursts in the records 
of the 1993 GALILEO-MARS OBSERVER-ULYSSES coincidence experiment is 
currently in progress \cite{UMOGcoinc}. However this analysis must be a rough 
one, as the structure of the signal in this 
strongly relativistic regime is essentially unknown and the use of data collected
in coincidence by several instruments is crucial.

The the paper is organized as follows. In Sec. \ref{sec;signal} we review 
the basic concepts and formula (partially to fix notation) regarding
the gravitational wave emission from binaries. 
In Sec. \ref{sec;doppler} we introduce the signal $y(t)$ produced
at the output of a Doppler detector 
by a gravitational wave train $h(t)$: in the Fourier space, $\tilde{y}(f)$ is the
product of $\tilde{h}(f)$ and
the Doppler {\it three-pulse response function} $\tilde{r}_{\theta}(f)$; 
we review the main properties of $\tilde{r}_{\theta}(f)$ and in particular we 
stress its different structure in the low ($f \simlt 1/T$) and high ($f \simgt 1/T$)
frequency regimes. Sec. \ref{sec;sens} contains the main results of the paper:
a thorough analysis of the signal-to-noise ratio 
produced by in-spiral signals in Doppler experiments as a function of the 
parameters that characterize the source and the instrument: 
the chirp mass $\Mc$ (and the mass ratio), the 
instantaneous emission frequency of the signal $f_b$ when the instrument is "turned on", 
the duration of the observation $T_1$, the structure of the instrument
noise spectral density $S_n(f)$, the location $\theta$ of the source in the sky with
respect to the detector arm
and the round-trip light-time $T$. This analysis is carried out in 
full generality and can be applied to any Doppler experiment.
We define the {\it ideal} SNR, $\rho_{\rm id}$, as that corresponding to 
unlimited bandwidth, 
$\theta = \pi/2$ and flat noise spectrum:
$\rho_{\rm id}$ depends only on $\Mc\,,T$, $D$ and the noise spectral level $S_0$. 
Of course, in a real experiment several factors -- among which the
finite observation time, the signal bandwidth, the frequency-dependent
noise spectral density and the position of the source in the sky -- contribute
to degrade the SNR. Indeed, 
we define the {\it sensitivity function} $\Upsilon$ as the actual SNR
with respect to its ideal value. $\Upsilon$ is extensively studied
in Sec. \ref{subsec;sensf}. In particular we
show that the maximum value of $\Upsilon$ is achieved when $0.1/T \simlt
f_b \simlt 1/T$ and the binary chirp mass is close to what we call
its critical value $\Mc_c \propto T_1^{-3/5}\,f_b^{-8/5}$.
In Sec.  \ref{subsec;senscass} we study the CASSINI's sensitivity (and compare it
with that of ULYSSES), with emphasis on possible sources in the Virgo Cluster and the Galactic
Center. We show that the maximum reachable distance is $\sim 600$ Mpc;
binaries close to the final coalescence with $10^6\,\Ms \simlt \Mc
\simlt 10^9\,\Ms$ and comparable masses would be detectable in the Virgo cluster with
SNR up to 30; in galactic searches, CASSINI would be able to pick up (possibly in
all three data sets) signals from the in-spiral of a secondary black hole of
mass $M_2\simgt 50\,\Ms$ onto the central one 
$M_1\simeq 2\times 10^6\,\Ms$. Finally, Sec. V contains a discussion
about the probability of success of Doppler
experiments based on simple conventional astrophysical scenarios and our conclusions.

We take units in which $c=G=1$, so that velocities are dimensionless and 
$10^6\,\Ms \simeq  4.926 \,{\rm sec}$.

\section{THE EXPECTED SIGNAL}
\label{sec;signal}

\subsection{The chirped precursor}
\label{subsec;chirp}

We consider a binary system of two compact objects of mass
 $M_1$ and $M_2$; $M \equiv M_1 + M_2$ and $\mu \equiv
M_1 M_2/M$ are the total and reduced mass, respectively; the
orbital parameters evolve
secularly due to the loss of energy and angular momentum by emission
of gravitational waves. 

We shall assume that, when the gravitational signal
enters the sensitivity window of the detector, the radiation reaction
has already circularized the orbit. Under this assumption,
in the Newtonian quadrupolar approximation, the frequency $f$ 
of the emitted gravitational wave is twice the orbital frequency of
the binary, namely
\be 
f^2(t) = \frac{M}{\pi^2r^3(t)}\,,
\ee 
where $r(t)$ is the orbital separation. In the 
Newtonian model, $f(t)$ evolves according to \cite{PNapprox}
\be 
\dot f(t) = \frac{96}{5}\pi^{8/3}\Mc^{5/3}f^{11/3}(t)\,,
\label{fdot}
\ee 
where the dot indicates the time derivative and
\be  
\Mc \equiv \mu^{3/5} M^{2/5} = \eta^{3/5}M\,\quad\quad (\eta\equiv\frac{\mu}{M}
\le 1/4)
\label{cm} 
\ee 
is the so-called {\em chirp mass}; in this approximation $\Mc$ is the only 
{\it dynamical} parameter that 
regulates the evolution of the frequency and the amplitude
of the wave.

Let us consider a signal emitted by a binary at a luminosity distance $D$
and let $\iota$ be the angle between the wave propagation
direction and the orbital angular momentum
${\gras L}$. The strain at the detector reads
\ba
h(t) & = & h_+(t)\cos 2\varphi + h_{\times}(t) \sin 2\varphi \nonumber\\ 
& = & A(t)\,\cos{\left [2 \pi \int^t dt'f(t') +\varphi_0\right]}\,,
\label{ht} 
\ea 
where $ h_+(t)$ and $h_{\times}(t)$ are the two independent
polarization states \cite{MTW}, and $\varphi$  a polarization angle;
\ba
A(t) & = & 2\, Q(\iota, \varphi)
\frac{\Mc^{5/3}}{D} [\pi f(t)]^{2/3}\nonumber\\
& = & 1.3\times 10^{-15}\,Q(\iota, \varphi)\, \left(\frac{D}{{\rm Mpc}}\right)^{-1}
\left(\frac{f}{10^{-4}{\rm Hz}}\right)^{2/3}
\left(\frac{\Mc}{10^6\Ms}\right)^{5/3}\,,
\label{ampt}
\ea
is the amplitude of the signal and 
\be
\varphi_0 = {\rm tan}^{-1}\left[\frac{2\cos\iota \sin(2\varphi)}{\cos(2\varphi)
(1 +\cos^2\iota)}\right]
\label{polph}
\ee
gives the polarization phase. The quantity
\be  
Q^2(\iota, \varphi) = [\cos^2(2\varphi)
(1 +\cos^2\iota)^2 + 4\cos^2\iota \sin^2(2\varphi)]\,
\label{Qangl}
\ee
depends on the orientation of the binary and, for random values, 
$\m Q^2\M = 8/5$, where $\m \M$ stands for the average with respect
to $\iota$ and $\varphi$:
$Q(\iota, \varphi)$ ranges from 0 (for $\iota = \pi/2,
\varphi = \pi/4$) to 2 (for $\iota = 0$ or $\pi$) \cite{note_prec}. 

Integrating Eq. (\ref{fdot}), it is straightforward to derive the frequency
evolution of the wave:
\be 
f(t) = \frac{1}{8\pi}\left(\frac{5}{t_n}\right)^{3/8}\,\Mc^{-5/8}
\left(1-\frac{t}{t_n}\right)^{-3/8}\,, 
\label{ft}
\ee
and therefore the phase of the gravitational signal reads
\be 
\Phi(t) = 2 \pi \int_{t}^{t_n} dt'f(t')
= \Phi_n-2\left(\frac{t_n-t}{5\Mc}\right)^{5/8}
\label{faset}\,;
\ee
equivalently, 
\be t(f) = t_n - 5(8\pi f)^{-8/3 }\Mc^{-5/3}\,,
\label{tf}
\ee
\be 
\Phi(f) = \Phi_n-2(8\pi\Mc f)^{-5/3}\,.
\label{fasef}
\ee
From Eq. (\ref{fdot}), one can also derive the number of Newtonian wave cycles spent by a 
binary while it sweeps the relevant frequency interval
\be
{\cal N}(f) = \int_{f}^{\infty} \frac{f'}{{\dot f'}}\,df' = {\cal N}_n - 
\frac{1}{\pi} \,(8\,\pi\Mc\,f)^{-5/3}\,.
\ee
$t_n$, $\Phi_n$ and ${\cal N}_n$ are integration constants, defined as the
values that $t(f)$, $\Phi(f)$ and ${\cal N}(f)$ formally take when $f=\infty\,(r=0)$.
Of course, the signal must be cut off when the in-spiral phase ends and the 
merger begins. The fully relativistic two-body problem 
is still unsolved; it is mainly investigated using post-Newtonian approximations
(see \cite{will94,blanchet96} and references
therein) and numerical techniques (see \cite{Seidel97,GranC,Pullin} and
references therein). For simplicity we will neglect the
radiation coming from frequencies higher than 
\be
f_{\rm isco} = \frac{1}{6^{3/2}\,\pi\,M} 
\simeq 1.9\times 10^{-3}\,\left(4\eta\right)^{3/5}
\,\left(\frac{\Mc}{10^6\,\Ms}\right)^{-1}\,{\rm Hz}\,,
\label{fisco}
\ee
corresponding
to the {\it innermost stable circular orbit} $r_{\rm isco} = 6\,M$ for a 
test mass in the Schwarzschild field of a mass $M$. 
For $r < r_{\rm isco}$ (and velocity larger than 
$v_{\rm isco} \simeq 0.408$) orbits are unstable. For general black 
holes and mass ratios we can write $r_{\rm isco} = 6\,k\,M$, with 
$k \approx 1$ (several methods have been used
to estimate the exact value of $k$, but a satisfactory answer still lacks;
see \cite{BD,KWWa,KWWb,Cook} for further details). In the Newtonian approximation we 
assume that $f_{\rm isco} = f(t_c)$ 
separates, at the time $t_c$, the in-spiral phase from the broad band plunge-in. 
The major feature of the relativistic corrections is, therefore, to {\em anticipate} 
the final plunge. 

If an experiment takes place in the time interval from $t = 0$ to $t = T_1$,
we define
\be
f_b \equiv f(t = 0)
\label{fb}
\ee 
as the frequency at which the signal enters the record at $t = 0$; 
the frequency at the end, after the observation time $T_1$, is
\be 
f_e \equiv f(t = T_1) = \min\left[
f_b\left(1-\frac{T_1}{t_n}\right)^{-3/8},\,f_{\rm isco}\right]\,. 
\label{fe} 
\ee 
If $f_b (1-T_1/t_n)^{-3/8} > f_{\rm isco}$ the record includes
the merger signal and is of class IV; otherwise  we have
class III signals. The Newtonian time to coalescence for a system 
radiating at frequency $f_b$ at the beginning
of the record is given by:
\be 
\tau_n = 5\, (8 \pi f_b)^{-8/3}\, \Mc^{-5/3}\,.
\label{tn}
\ee
It is also useful to note that the time for a signal to sweep from the
frequency $f_b$ to the beginning of the final coalescence, that we suppose
to occur at $f = f_{\rm isco}$, is
\be
\tau_c = 5\, (8 \pi)^{-8/3}\, \Mc^{-5/3}\,
\left(f_b^{-8/3} - f_{\rm isco}^{-8/3}\right)\,.
\label{Tc}
\ee
We define
\be
f_B \equiv f_{\rm isco} \left(1+\frac{T_1}
{5 (8 \pi f_{\rm isco})^{-8/3}\,\Mc^{-5/3}}\right)^{-3/8}
\label{fbmax}
\ee
as the initial frequency of a gravitational wave whose final frequency is $f_{\rm isco}$;
it depends on $\Mc$, $\eta$  and $T_1$ and corresponds to the largest 
initial frequency that produces class III signals. 

\begin{figure}
\centerline{\mbox{\epsfysize=15.cm
\epsffile{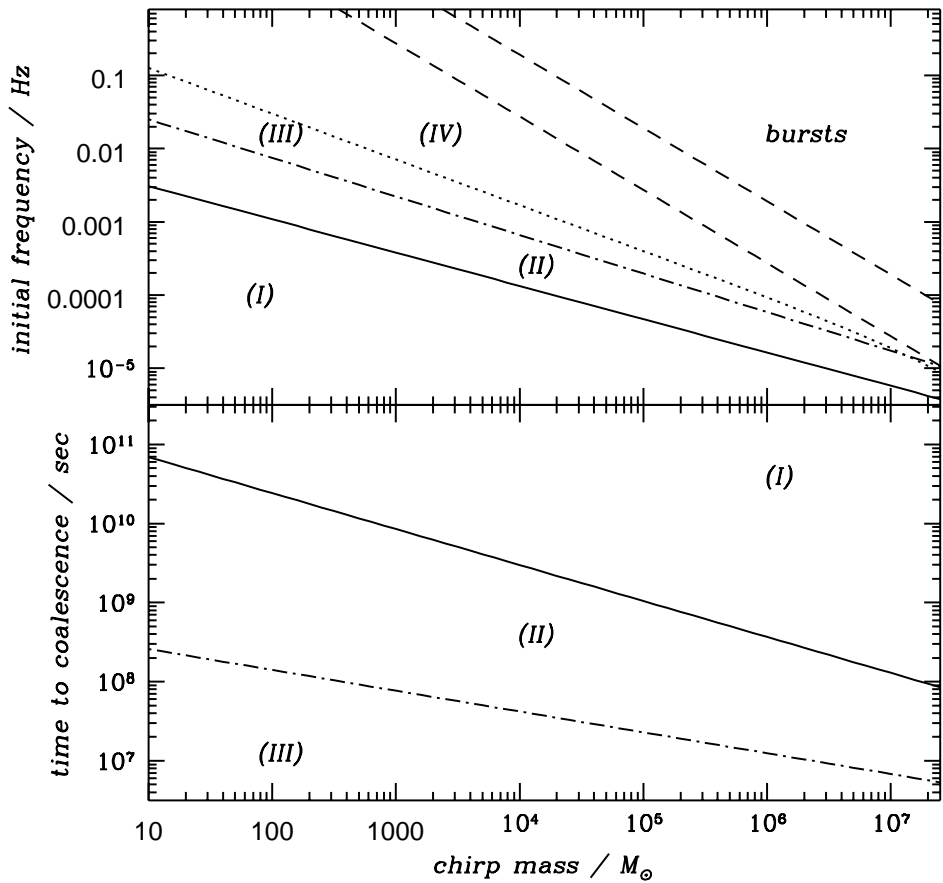}}}
\caption{\label{fig;freqreg}
Classes of coalescing binary signals for one of the three CASSINI's experiments
($T_1 = 40\, {\rm days}$). 
The upper panel shows the partition of the plane
$(\Mc,f_b)$ in periodic (I),  linear (II) and non-linear region.
The dotted line divides non-linear chirps in 
signals of class III and IV.
The dashed lines indicate the boundary between signals of
class IV and bursts -- 
{\rm i.e.} $f_{\rm isco}$ as a function of
$\Mc$ -- for $M_2/M_1 = 0.01$ (lower line) and $1$
(upper line). The lower panel shows the  Newtonian times to
coalescence $\tau_n$, Eq. (\ref{classtau}), computed at the transition frequency
(see upper panel) for each class of signals.
}
\end{figure}

\begin{figure}
\centerline{\mbox{\epsfysize=15.cm
\epsffile{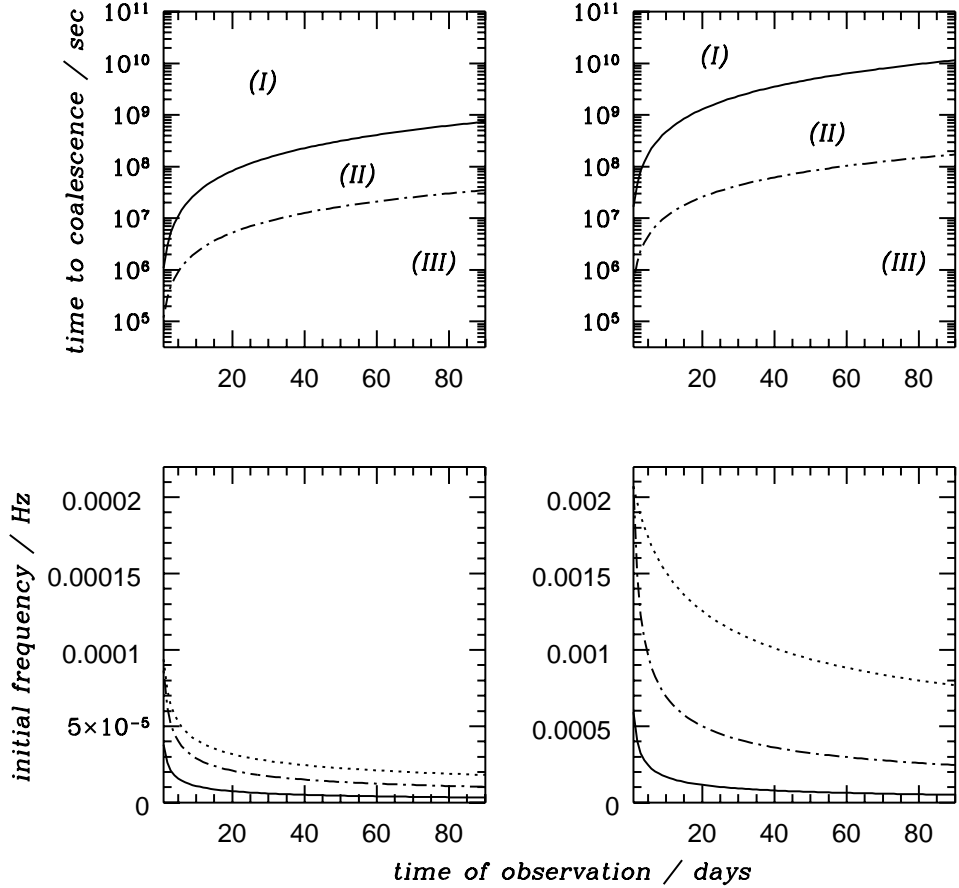}}}
\caption{\label{fig;fT1}
To show how the partition of signals into different classes depend upon the observation 
time $T_1$ we consider two binary systems: $M_1 = M_2 = 10^7 \,\Ms $ 
(left panels) and $M_1 = 2\times 10^6 \Ms\,, M_2 = 1000\, \Ms$ 
(right panels). The Newtonian  time to coalescence $\tau_n$, Eq. 
(\ref{classtau}),
and the initial frequency $f_b$, Eq. (\ref{classfreq}), are used to separate the classes: 
continuous line between I and II, dashed-dotted line between II and III, 
dotted line between III and IV. The frequency scale has been chosen so
that $f_{\rm isco}$ corresponds to the top of the panel.
}
\end{figure}

On the base of the model of the waveform we can now formalize the 
distinction of the signals in classes that we have introduced in Sec. I. 
Being based upon the actual record, it is particularly useful for data
analysis. It refers to the frequency resolution $1/T_1$ of
the experiment and will allow us to divide the two-dimensional 
parameter space $(\Mc,f_b)$, where the in-spiral signal is defined, into
a {\it periodic region}, a {\it linear region} and a {\it general chirp
region} (of class III and IV, depending whether the final burst is
within the record or not). This distinction is justified by the 
great simplicity and low computational load
of the data analysis when the frequency
is constant (class I) or varies linearly with time (class II); these two
cases are extensively discussed, and applied to the data of ULYSSES'
second opposition, in \cite{pap1}. When
\be
T_1^2\dot{f}(t) = \frac{96}{5}\,\pi^{8/3}T_1^2\Mc^{5/3}f^{11/3} < 1
\quad\quad ({\rm class\,\, I})
\label{periodic}
\ee 
we have periodic signals; when 
\be
T_1^3\frac{\ddot{f}(t)}{2} = 
\frac{16896}{25}\,\pi^{16/3}\,T_1^3\,\Mc^{10/3} f^{19/3} < 1
\quad\quad {\rm and} \quad\quad T_1^2\dot{f}(t) > 1
\quad\quad ({\rm class\,\, II})
\label{linear}
\ee
linear signals;
\be
\frac{T_1^3\,\ddot{f}(t)}{2} > 1 
\quad\quad {\rm and} \quad\quad f_b < f_{\rm isco}
\quad\quad ({\rm class\,\, III/IV})
\label{general}
\ee
correspond to general chirps \cite{note_class}.
Referring to a chirp mass $\Mc$ and observation time $T_1$, 
the initial frequency $f_b$ that separates out 
the four regimes reads:
\begin{eqnarray}
f_b \simeq
\left\{ 
\begin{array}{ll}
5.1\times 10^{-5}
\left(\frac{T_1}{10^6{\rm sec}}\right)^{-6/11}
\left(\frac{\Mc}{10^6\Ms}\right)^{-5/11}\,{\rm Hz} & 
(I-II)
\\
 5.4\times 10^{-5}
\left(\frac{T_1}{10^6{\rm sec}}\right)^{-9/19}
\left(\frac{\Mc}{10^6\Ms}\right)^{-10/19}\,{\rm Hz} &
(II-III)
\\
1.9\times 10^{-3}\,\left(4\eta\right)^{3/5}\,
\left(\frac{\Mc}{10^6\,\Ms}\right)^{-1}\,
\left[1 + 8.7\times 10^2 \,\left(4\eta\right)^{8/5}
\left(\frac{T_1}{10^6\,{\rm sec}}\right)
\left(\frac{\Mc}{10^6\,\Ms}\right)^{-1}\right]^{-8/3}\,{\rm Hz} &
(III-IV)
\end{array} 
\right.\,.
\label{classfreq} 
\end{eqnarray} 
The partition of the plane $(\Mc,f_b)$ in classes for a single CASSINI's
experiment ($T_1 = 40\,{\rm days}$) is given in Fig. \ref{fig;freqreg}. 
The values of $\tau_n$ that correspond to the transition between two
adjacent classes read
\begin{eqnarray}
\tau_n \simeq
\left\{ 
\begin{array}{ll}
6.1\times 10^7\,\left(\frac{\Mc}{10^6\,\Ms}\right)^{-5/11}\,
\left(\frac{T_1}{10^6{\rm sec}}\right)^{16/11}\,{\rm sec} &  
\quad (I-II)
\\
2.6\times 10^6\,\left(\frac{\Mc}{10^6\,\Ms}\right)^{-5/19}
\,\left(\frac{T_1}{10^6{\rm sec}}\right)^{24/19}\,{\rm sec}
 &
\quad (II-III)
\\
1.1\times 10^3\,\left(\frac{\eta}{1/4}\right)^{-8/5}\,
\left(\frac{\Mc}{10^6\,\Ms}\right)\,{\rm sec} &
\quad (III-IV)
\end{array} 
\right.\,.
\label{classtau}
\end{eqnarray}
As we have
mentioned in the Introduction, class I to IV correspond to gravitational
wave emission increasingly closer to the final merger and,
therefore, to stronger signals.
In Fig. \ref{fig;fT1} we show
the frequency $f_b$ and the characteristic time $\tau_n$ that
separate the four classes as function of the observation
time. For typical present and future data 
sets (see Table
\ref{tab;doppexp}) and reasonable sources, most signals would show up 
in class III or IV.

We introduce now the Fourier transform $\tilde h(f)$ of the real function $h(t)$: 
\be  
\tilde h(f) = \int_{-\infty}^{\infty} e^{- 2\pi i ft} h(t) \,dt\, =
\tilde h^*(-f)
\label{four} 
\ee 
and hereafter we will work in the more convenient Fourier space. 
When the frequency $f(t)$ and the amplitude $A(t)$ (Eqs. (\ref{fdot}) and 
(\ref{ampt})) vary over a time scale much longer than $1/f(t)$ it is 
convenient to use the {\em stationary phase approximation} in order to
compute the integral (\ref{four}). In our case
\be 
\frac{1}{f(t)}\,\frac{d}{dt}\,\ln[A(t)]\propto \frac{1}{f(t)}\,\frac{d}{dt}\,
\ln[f(t)]
\propto \left[\Mc f(t)\right]^{5/3}\,,
\ee
where we have neglected numerical coefficients of proportionality. 
The evaluation of the actual error involved in this approximation 
is a delicate matter; however, in the limit
\be
\left[\Mc f(t)\right]^{5/3} \ll 1 \,,
\label{spalim}
\ee
we expect this approximation to be satisfactory for the assessment of the 
signal-to-noise ratio. Note also that
\be 
\left[\Mc f(t)\right]^{5/3} < (\Mc f_{\rm isco})^{5/3} \propto 
\left (\frac{\Mc }{M}\right )^{5/3} \propto \eta\,, 
\ee
so that the limit (\ref{spalim}) is formally fulfilled for a small mass ratio. Of course, as usual, 
the approximation will be fearlessly pushed to the limits of validity.

In the stationary phase approximation to the integral (\ref{four}), 
for a frequency $f$ only a small interval $\delta t$ contributes around the 
time $t(f)$ where 
\be 
f(t) = f\,. 
\ee
Note the slight inconsistency of notation: $f$ is now the independent variable in 
the Fourier space, while earlier we have introduced $f(t)$ as a function of time. 
The time interval is of order
\be
\delta t =[\dot{f}(t)]^{-1/2} = \left(\frac{5}{96}\right)^{1/2}
\,\pi^{-4/3}\,\Mc^{-5/6}\,f^{-11/6}(t),
\ee 
corresponding to a local bandwidth $\delta f = 1/\delta t$. A finite record 
in $(0,T_1)$  will pick up only the part of the spectrum between $f_b$
and $f_e$; since dropping the data outside
$(0,T_1)$ is equivalent to convolving with $\sin (fT_1)/f$, one
wonders how this can kill the Fourier components outside
$(f_b, f_e)$. The answer is that outside the frequency range defined
by $f = f(t) ,\,(0\le t\le T_1)$, the phase of the
Fourier integral (\ref{four}) has fast oscillations. 
The Fourier amplitude is then of order 
\be
A \delta t \approx \frac{\Mc^{5/6}}{D}\,f^{-7/6}\,.
\ee
The full expression of the Fourier transform of $h(t)$,
Eq. (\ref{ht}), computed according to the stationary phase approximation
is \cite{CF}:
\begin{eqnarray}
\hf(f) =  
\Biggl\{ 
\begin{array}{ll}
{\cal A} f^{-7/6} e^{i \Psi(f)}
&  \quad f \le f_{\rm isco}
\\
0 & \quad f > f_{\rm isco}
\end{array} 
\Biggr.\,, 
\label{stpha}
\end{eqnarray}
where 
\be 
{\cal A} =  \left(\frac{5}{6}\right)^{1/2} \frac{1}{4\pi^{2/3}} Q(\iota, \phi)\frac{\Mc^{5/6}}{D} 
\label{fourha} 
\ee
is the amplitude and 
\be
\Psi(f)  = 2 \pi f t_n - \Phi_n - \frac{\pi}{4} + \frac{3}{4} (8 \pi \Mc f)^{-5/3} 
\label{fourhph}  
\ee 
the phase.

The existence of an analytic expression of the in-spiral signal in the Fourier space justifies 
its use. As compared with the analysis in the time domain, it offers two main
advantages: the instrumental response function,
see Eqs. (\ref{yt})--(\ref{rf}), is a simple 
multiplicative factor and we can easily deal with coloured noise 
(see Sections \ref{sec;doppler} and \ref{sec;sens}).

\subsection{The final burst}
\label{subsec;burst}

The theoretical knowledge of the signal emitted during the 
merger of two black holes is at present quite poor. It involves the
computation of the solutions of the Einstein equation in highly non linear
regimes and this issue is currently tackled by several groups mainly
by means of intensive numerical computation on state-of-the art
machines (see \cite{Seidel97,GranC,Pullin} and references therein).
Here we present a simple model for the energy spectral density $dE/df$
emitted during the merger, following the approach presented in \cite{FH}; although
quite naive and crude, it is reasonable for an order of magnitude comparison
of the sensitivity of Doppler experiments to in-spiral and merger
signals, which shall be carried out in Sec. \ref{sec;sens}.

The wave emitted during the final stages of
the life of a binary system is spread over a wide band. For sake of simplicity,
and lack of more detailed knowledge, we assume it to be confined between the final 
frequency of the in-spiral, {\it i.e.} $f_{\rm isco}$, and 
the characteristic frequency $f_{\rm qnm}$ at which the ring-down emission sets in. We will assume 
\be
f_{\rm qnm} \simeq \frac{1}{2\,\pi\, M} = 
3.2\times 10^{-2}\,\left(\frac{M}{10^6\,\Ms}\right)^{-1}\,{\rm Hz}
\label{fqnm}
\ee
that roughly corresponds to the frequency of the quasi-normal oscillation for the 
$l = m = 2$ mode of a  Kerr black hole with dimensionless spin parameter $\simeq 0.98$.
This choice of $f_{\rm qnm}$ can be justified as follows: the $l = m = 2$ mode,
very likely to be excited at the end of the black hole plunge-in, is
that with the longest damping time and the resulting massive black hole will quite
probably carry a large intrinsic angular momentum.

The total energy $E_{\rm merg}$
radiated during the merger phase, and confined into the frequency interval
$(f_{\rm isco},f_{\rm qnm})$, is proportional to $\mu^2/M$ \cite{Smarr} and can be written as
\be
E_{\rm merg} = \epsilon_{\rm merg}\, F\left(\frac{\mu}{M}\right)\, M\,,
\label{Em}
\ee
where $\epsilon_{\rm merg}$ denotes the efficiency of the process and $F(\mu/M) = (4\mu/M)^2$; 
of course, a precise estimate of $\epsilon_{\rm merg}$ would require a full numerical
calculation, but a reasonable order of magnitude guess
yields $\epsilon_{\rm merg} \sim 0.1$ (see \cite{FH} and references therein). Lacking any
evidence about the structure of the energy spectrum, we will make the simple
assumption that $E_{\rm merg}$ is uniformly distributed over the band
$(f_{\rm qnm},f_{\rm isco})$, so that
\begin{eqnarray}
\frac{dE_{\rm merg}}{df} = 
\Biggl\{ 
\begin{array}{ll}
\frac{E_{\rm merg}}{(f_{\rm qnm} - f_{\rm isco})} & \quad f_{\rm isco}  \le f \le f_{\rm qnm}\\
0 & \quad f < f_{\rm isco}\,\,\, {\rm and}\,\,\, f > f_{\rm qnm}
\end{array} 
\Biggr.\,. 
\label{dEmdf}
\end{eqnarray}
We recall here that the energy spectrum of the radiation emitted during the
in-spiral phase is given by
\begin{eqnarray}
\frac{dE_{\rm insp}}{df} = 
\Biggl\{ 
\begin{array}{ll}
\frac{\pi^{2/3}}{3}\,\Mc^{5/3}\,f^{-1/3} & \quad f \le f_{\rm isco}  \\
0 & \quad  f > f_{\rm isco}
\end{array} 
\Biggr.\,. 
\label{dEinspdf}
\end{eqnarray}
Eqs. (\ref{dEmdf}) and (\ref{dEinspdf}) will enable us to compare the detectability
of coalescing binaries during the in-spiral and merger phase with Doppler experiments, as the 
optimal signal-to-noise ratio is simply related to $dE/df$, see Eq. (\ref{rhoE}); this
analysis is carried out in Sec. \ref{sec;sens}.

\section{DOPPLER RESPONSE TO IN-SPIRAL SIGNALS}
\label{sec;doppler}

In a Doppler experiment, the Earth and a spacecraft are used as the 
end-points of a gravitational wave detector \cite{EW}. A radio link
is transmitted from the Earth to the spacecraft, 
coherently transponded and sent back to the Earth, where, at the time
$t$, its frequency 
is measured with great accuracy; comparing the emitted and received 
frequency -- $\nu_0$ and $\nu(t)$, respectively --
one determines the Doppler shift $(\nu(t)-\nu_0)/\nu_0$ 
as a function of time. The contribution to this effect produced by a gravitational
wave $h(t)$ is usually called the {\it Doppler signal} $s(t)$ and reads \cite{EW}
\be  
s(t) = \int_{-\infty}^{\infty} dt'r_{\theta}(t- t') h(t')\,; 
\label{yt}
\ee 
\be
r_{\theta} (t) =  \frac{\cos \theta-1}{2} \delta(t) -\cos \theta\,
\delta \left (t - \frac{\cos \theta +1}{2}T \right)
+ \frac{1 +\cos \theta}{2} \delta(t - T)\,
\label{rt} 
\ee
is the characteristic {\it three-pulse response} of the instrument, as the 
incoming signal $h(t)$ is repeated in the detector output at three different
times (and with different amplitude): $t$, $t -(\cos \theta +1)T/2$ and $t-T$; 
$r_{\theta} (t)$
depends on the angle $\theta$ between the spacecraft and the source
and the round-trip light-time $T = 2L$ out to the distance $L$
of the probe. In the frequency domain the Doppler signal (\ref{yt})
takes a very simple form:
\be
\tilde s(f) = \rf_{\theta}(f) \hf(f)\,,
\label{yf}
\ee
where $\hf(f)$ is given by Eq. (\ref{stpha}) and the Fourier transform
of $r_{\theta}(t)$ reads
\be
\tilde r_{\theta}(f) =  \frac{\cos \theta -1}{2} - \cos \theta
\exp{[\pi i(1 +\cos \theta)f T]} + \frac{1 +\cos
\theta}{2} \exp[2 \pi i f T]\,.
\label{rf}  
\ee
Since the frequency $f$ appears multiplied by the round-trip
light-time $T$, we introduce the
dimensionless variable
\be
\xi \equiv f T\,.
\label{fxi}
\ee

\begin{figure}
\centerline{\mbox{\epsfysize=15.cm
\epsffile{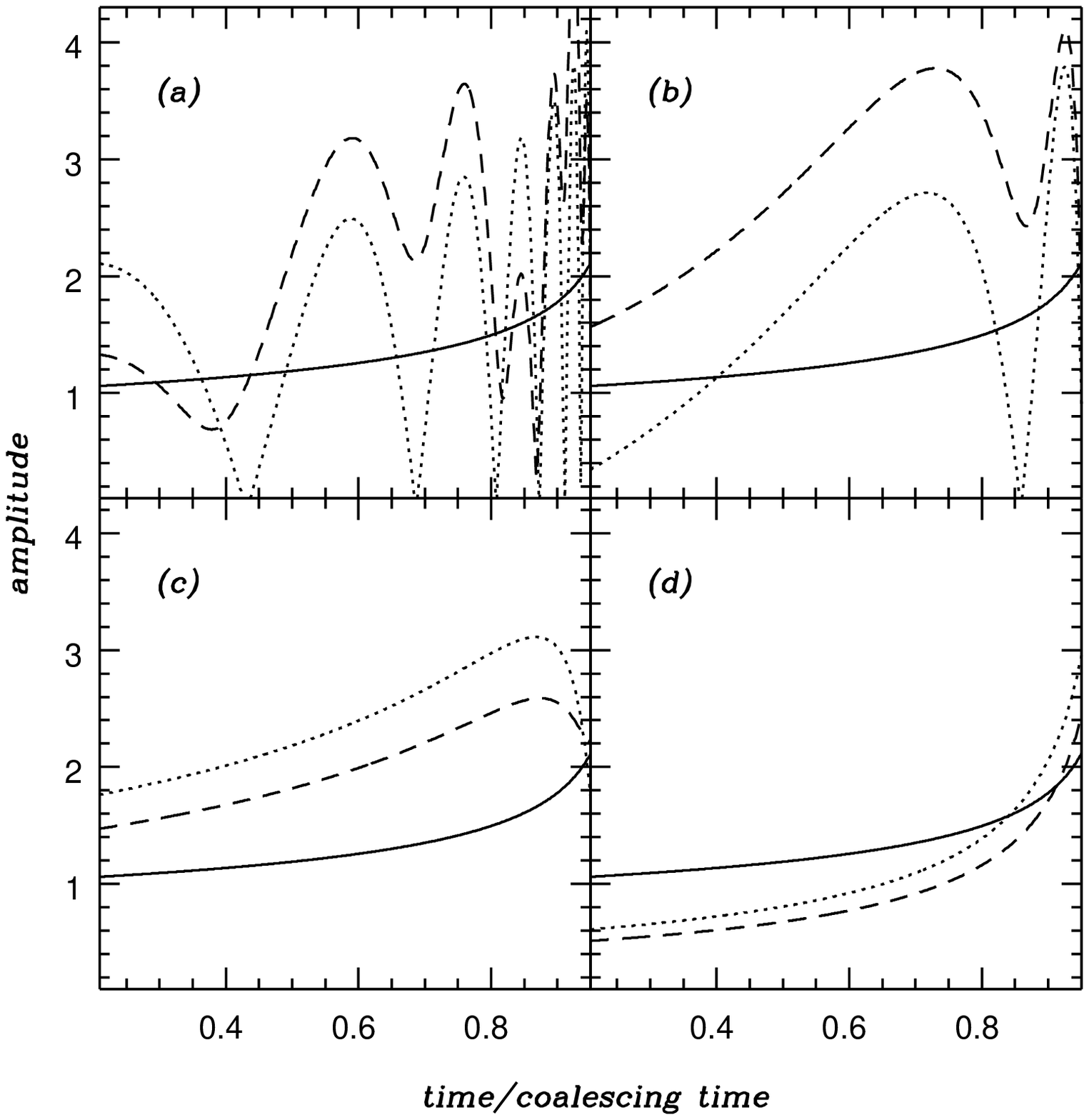}}}
\caption{\label{fig;ampdopp}
The effect of the Doppler filter $r_{\theta}$ on in-spiral signals. 
The plots show a comparison between the chirp amplitude 
$A(t)$, Eq. (\ref{ampt}),  (solid line) and the Doppler amplitude
for different values of the angle $\theta$ ($\cos\theta = 0$: dashed-line;
$\cos\theta = 0.4$: doted-line) and of the chirp mass: (a) $\Mc = 10^5\,\Ms$,
(b) $\Mc = 10^6\,\Ms$, (c) $\Mc = 10^7\,\Ms$, (d) $\Mc = 10^8\,\Ms$
($\eta = 1/4$ for all systems).
The chirp amplitudes are normalized to $A(t=0) = 1$; the round-trip 
light-time is $T = 4000 $ sec. Note the increasing
modulation induced by the Doppler response as the final merger is 
approached; this modulation disappears as $\Mc/T$ increases because 
the filter degenerates in a multiplicative factor (cfr. Eq. (\ref{rfll1})
and Fig \ref{fig;doppresp}).
}
\end{figure}

In agreement with the
Weak Equivalence Principle, the response of the instrument goes to zero in the
low frequency limit $\xi\ll 1$ (this is the regime in which 
ground and space-based interferometers work):
\be 
\rf_{\theta}(\xi) = i\pi \xi \sin^2{\theta} 
- \frac{1}{2}\pi^2\xi^2\sin^2\theta (2 + \cos\theta) + O(\xi^3)\,.
\label{rfll1}
\ee 
To lowest order the angle $\theta$ only affects the coefficient. As $\xi$ increases, 
$\rf_{\theta}(\xi)$ develops modulations in amplitude and, more importantly, in
phase, that depend on ${\theta}$. The beats produced by the superposition of the signals at the
times $t$, $t -(\cos \theta +1)T/2$ and $t-T$ result in minima and
maxima of the amplitude superimposed to the general increasing
trend $A(t) \propto f^{2/3}(t)$, see Eq. (\ref{ampt}), as shown in Fig \ref{fig;ampdopp}; 
they strongly depend on the angle $\theta$ and provide a
sensitive signature for its determination.

The transversal character of gravitational waves is reflected in the 
forward $(|\theta| \ll 1)$ and backward $(|\theta'| = |\pi - \theta| \ll 1)$
limits, when the wave travels in a direction almost parallel to the 
Earth-spacecraft line:
\be 
\rf_{\theta}(\xi) = \frac{\theta^2}{4}\left[(1 + 2\pi i\xi)e^{2\pi i \xi} - 1
\right] +O(\theta^4)\quad\quad (|\theta| \ll 1)\,,
\label{rflim1} 
\ee
\be
\rf_{\theta}(\xi) = \frac{\theta'^2}{4}\left [ e^{2\pi i \xi} - 1 + 2\pi
\xi\right] +O(\theta'^4)\quad\quad (|\theta'| = |\pi - \theta| \ll 1)\,;  
\label{rflim2}  
\ee 
both tend to zero; they have the same limiting amplitude, but different 
phases. Note that the two limits are not uniform and are violated at high frequency
$(\xi\gg 1)$.

\begin{figure}
\centerline{\mbox{\epsfysize=15.cm
\epsffile{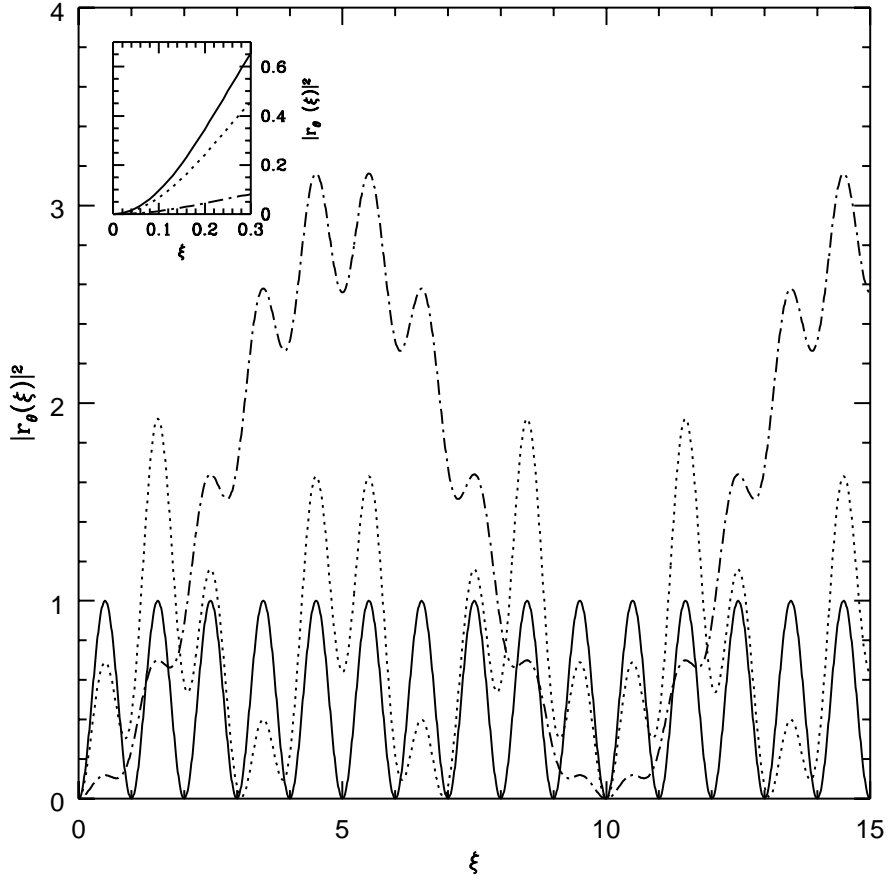}}}
\caption{\label{fig;doppresp}
The square modulus of the Doppler filter
$|\tilde{r}_{\theta}(\xi)|^2$, Eq. (\ref{rf2}), as a function of
$\xi\equiv f\,T$ for selected values of the angle $\theta$ (solid line:
$\cos\theta = 0$; dotted-line: $\cos\theta = 0.4$;
dotted-dashed line: $\cos\theta = 0.8$). The small box zooms the
behavior of $|\rf_{\theta}(\xi)|^2$ for $\xi\ll 1$; note that,
only when  $\xi$ is small, $|\tilde{r}_{\theta}(\xi)|^2$ is a monotonic decreasing 
function of $\cos\theta$, which is not true for $\xi\simgt 1$.
}
\end{figure}

For generic values of $\theta$ it is convenient to use the squared modulus
of (\ref{rf}):
\ba  
|\rf_{\theta}(\xi)|^2 & = & \frac{3 \cos^2 \theta +1}{2} + 
\frac{\cos^2 \theta -1}{2} \cos(2\pi \xi) \nonumber\\
& & - \cos \theta(\cos \theta +1)\cos[(\cos \theta -1)\pi \xi ] \nonumber\\ 
& & - \cos \theta (\cos \theta -1)\cos[(\cos\theta+1) \pi \xi ]\,,
\label{rf2}
\ea 
which, in the limit $\xi\ll 1$, reads
\be  
|\rf_{\theta}(\xi)|^2 = \pi^2\xi^2\sin^4\theta
\left[1 - \frac{1}{3} \left(1 - \frac{1}{4} \cos^2\theta \right)  \pi^2\xi^2
\right] + O(\xi^5)\,.
\label{rf2ll1}
\ee
$|\rf_{\theta}(\xi)|^2$, which is shown in Fig  \ref{fig;doppresp}, is
an even function of $\cos \theta$, sum of four functions periodic 
in $\xi$, with frequencies 0,
$1, (1-\cos\theta)/2$ and $(1 +\cos\theta)/2$,  respectively.
Its average is $(3 \cos^2 \theta +1)/2$. As example, for 
$\theta \rightarrow \pi/2$ the response reads
\ba  
|\rf_{\theta}(\xi)|^2 & = & 
\sin^2(\pi\xi) +\bigl[(1 - \cos(\pi\xi))^2\nonumber\\ 
& -  &  2\pi\xi \sin(\pi\xi)\bigr]
\bigl(\theta-\frac{\pi}{2}\bigr)^2 +O\bigl[\bigl(\theta-\frac{\pi}{2}\bigr)^4
\bigr]
\ea
and it has a maximum for $\xi = 1/3$. 
For large frequencies it
develops more and more lobes with the angular scale $1/\xi$ 
(see Fig. \ref{fig;doppresp}). Averaging $|\rf_{\theta}(\xi)|^2$ over the direction $\theta$ one
obtains the mean square response of the detector:
\ba
\overline{|\rf_{\theta}(\xi)|^2} & = & \frac{1}{2}\int_{-1}^1
|\rf_{\theta}(\xi)|^2\,d(\cos\theta) \nonumber\\
& = & \frac{\pi^2\xi^2-3}{\pi^2\xi^2} - \frac{\pi^2\xi^2+3}{3\pi^2\xi^2}
\cos(2\pi\xi)+\frac{2}{\pi^3\xi^3}\sin(2\pi\xi)\,,
\label{rfaver}
\ea
which, in the low and high frequency limit, reads:
\begin{eqnarray}
\overline{|\rf_{\theta}(\xi)|^2} =  
\Biggl\{ 
\begin{array}{ll}
\frac{8}{15}\pi^2\xi^2\left(1- \frac{29}{84}\pi^2\xi^2\right) + O(\xi^6)
&  \quad (\xi \ll 1)
\\
1 - \frac{1}{3} \cos (2\pi\xi) & \quad (\xi \gg 1) 
\end{array} 
\Biggr.\,. 
\label{rfaverlim}
\end{eqnarray}

During an experiment neither the angle $\theta$ nor the round-trip light-time 
$T$ are exactly constant, mainly due to the motion of the Earth; their
time-variation induces changes in the
phase of the filter of order $\pi \Delta\theta\,  fT$ and $\pi f \Delta T$. 
For the latter $\Delta T \approx 2\, v_{\oplus} T_1 /c \approx 2\times 10^{-4} T_1$ 
($v_{\oplus} \simeq 30$ km/sec is the orbital velocity of the Earth) and the effect 
is relevant when
\be
2\times 10^{-4} \pi f T_1 \approx 1\,,
\ee
while the angle $\theta$ affects the phase by a smaller amount. For CASSINI's experiments this occurs 
at about 5 mHz, which is above the most important sensitivity region of the 
instrument (see next Section). Indeed, we will consider constant both
$T$ and $\theta$.

\section{Sensitivity of Doppler experiments}
\label{sec;sens}

The Fourier output of the detector
\be 
{\tilde y}(f) = {\tilde s}(f) + {\tilde n}(f)
\label{yf1}
\ee
is the sum of the signal $ {\tilde s}(f)$  and the noise
${\tilde n}(f)$. The noise is 
assumed to be stationary, with zero mean and spectral power density
\be
\langle {\tilde n}^*(f) {\tilde n}(f') \rangle = \delta (f-f') S_n(f)\,.
\label{corrn}
\ee
We consider single-sided noise spectra $S_n(f)$, formally from $0$ to $+ \infty$,
and take into account the finite instrumental bandwidth with suitably 
large values at the boundaries of the sensitivity window. When $S_n(f)$ is white 
the corresponding {\it Allan deviation} \cite{allan}
\be 
\sigma_y = \sqrt{\frac{S_n}{\tau}} 
\label{sigmay}
\ee
is inversely proportional to the square root of the integration time 
$\tau$; for coloured spectra, see, e.g., \cite{Barnes}.
For simplicity we use a continuous spectrum $S_n(f)$, 
although, over an experiment of duration 
$T_1$ and with sampling time $\Delta_s$, the spectrum consists of
discrete Fourier modes spaced by $1/T_1$, from $1/T_1$ to
the Nyquist frequency $1/(2\Delta_s)$.
 
Since the shape of the signal in the in-spiral phase can be accurately 
predicted, the technique of {\em matched filtering} is particularly suitable
(see, e.g., \cite{Schutz89,Schutz97} and references therein). 
It can be shown that if the signal ${\tilde s}(f)$ is known to within a constant (and real) 
amplitude $h$, its least square estimator has a variance given by 
\be
\left(\frac{s}{\sigma_n}\right)^2 = \rho^2 =  4
\int_0^\infty \frac{|\tilde s(f)|^2}{S_n(f)}\,df \,.
\label{snropt} 
\ee
The quantity $\rho$ is appropriately called the {\it signal-to-noise ratio} and is related to the
energy spectrum of the radiation $dE/df$ by the relation
\be
\rho^2 \propto \frac{1}{D^2}\,\int_0^\infty \frac{|\rf_{\theta}(f)|^2}{f^2\,S_n(f)}\,
\frac{dE}{df}\,df \,.
\label{rhoE}
\ee

It is useful to introduce an ideal reference value of the SNR and 
to discuss its dependence on the main parameters.
From Eq. (\ref{rfaverlim}), we can approximate the instrumental filter 
$r_{\theta}$, as follows:
\ba 
|\rf_{\theta}(\xi)|^2 \sim
\left\{ 
\begin{array}{ll}
\xi^2 & \quad (\xi < 1)
\\ 
1 & \quad (\xi > 1) 
\end{array} 
\right.\,.
\label{rfasy} 
\ea
With a white noise spectrum $S_n = \sigma_y^2 \tau  = {\rm const.}$,  and taking into 
account the frequency dependence of the signal (see Eqs. (\ref{stpha}), (\ref{yf})
and (\ref{rfasy})), we have
\ba
\frac{|\tilde s(f)|^2}{S_n(f)} \propto 
\Biggl\{ 
\begin{array}{ll}
\frac{\xi^{-1/3}}{\sigma_y^2 \tau}  & \quad  (\xi < 1)\,\\
\frac{\xi^{-7/3}}{\sigma_y^2 \tau}   & \quad  (\xi > 1)
\end{array} 
\Biggr.\,. 
\label{snrasy}
\ea
This is a decreasing function of $\xi$, both sides are integrable, so that in 
this ideal case 
the bulk of the sensitivity comes from $f \sim 1/T$ (cfr. also
Fig \ref{fig;dupsdx}). 
We can define the {\em ideal} SNR by taking $\xi_b \equiv f_b\, T\ll 1$, $ \xi_e 
\equiv f_e\, T\gg 1$ and the
direction angle $\theta = \pi/2$, for which $|\rf_{\pi/2}(\xi)|^2 = 
\sin^2 (\pi\xi)$; from Eqs. (\ref{stpha}), (\ref{fourha}) and (\ref{snropt}),
averaging over the source angles $\iota, \varphi$, we get
\be 
\rho^2_{\rm id} = K \frac{T^{4/3} \Mc^{5/3}}{\sigma_y^2 \tau D^2}, 
\label{rhoid}
\ee
where 
\ba
K & \equiv & \frac{5}{24\pi^{4/3}}\, \frac{\m Q^2(\iota,\varphi)\M}
{T^{4/3}}
\,\int_{0}^{\infty}\,
f^{-7/3} |{\tilde r}_{\pi/2}(f)|^2\,df\nonumber\\
& = & \frac{2^{1/3}\,\pi}{3^{3/2}\,\Gamma\left(7/3\right)}
\simeq 0.64\,.
\label{effid}
\ea
The ideal {\it ken} -- the largest distance at which a source is ideally 
detectable at $\rho_{\rm id} = 1$ -- is
\be 
D_{\rm id} = K^{1/2}\, \frac{T^{2/3}\Mc^{5/6}}{\sigma_y \sqrt{\tau}}\,. 
\label{ken}
\ee
For the CASSINI's mission (see Table \ref{tab;doppexp}),
with $T = 10^4$ sec and 
$\sigma_y^2 \tau = 9\times 10^{-26}\,{\rm Hz}^{-1}$ , it has the value
\be  
D_{\rm id} \simeq 45\,
\left(\frac{\Mc}{10^6\Ms}\right)^{5/6}
\left(\frac{T}{10^4 {\rm sec}}\right)^{2/3}
\left(\frac{\sigma_y^2\tau}{9\times 10^{-26}\,{\rm Hz}^{-1}}\right)^{-1/2}
\,{\rm Mpc}\,.
\label{kencass}
\ee 
For ULYSSES the  
the ideal ken for $\Mc = 10^6 \Ms$, is $\simeq 3 \,{\rm Mpc}$; 
in the coincidence experiment, 
the values for GALILEO and MARS-OBSERVER are
$\simeq 0.5 \,{\rm Mpc}$ and $\simeq2 \,{\rm Mpc}$, respectively, where 
the values of $\sigma_y$, $\tau$ and $T$ have been chosen according to Table 
\ref{tab;doppexp}.

The ideal value (\ref{rhoid}) for the SNR neglects several degradation factors,
that are actually present in real experiments, including: 

\begin{enumerate}

\item The direction of the source may be unfavorable.

\item The frequency interval  $(f_b,f_e)$ swept by the signal is finite because 
the record has a finite length and the instrument has its own cutoffs.

\item The noise spectrum is generally red, which damages the SNR 
in the critical low frequency band.
\end{enumerate}

Only experiments can provide a safe estimate of the noise structure, 
especially when aiming at high sensitivities or wide frequency windows. 
Past Doppler data can be approximated with a simple analytical expression made up 
with power law contributions:

\be
S_n(f) = S_0 \sum_{j = 1}^3 \left(\frac{f}{f_j}\right)^{\alpha_j}\,,
\label{Sncol}
\ee
where 
\be
S_0 = S_n(1/T) = \sigma_y^2(T) T
\label{S0}
\ee
sets the spectral level; the frequencies $f_j$ and the exponents $\alpha_j$, fulfilling
\be
\sum_{j = 1}^3 (T\,f_j)^{-\alpha_j} = 1\,,
\label{Snorm}
\ee
characterize the different contributions to the noise. We can define the transition 
frequencies $f_{12}$ and $f_{23}$ 
as the intercept between two power approximations; for example, the transition 
frequency $f_{12}$ between region 1 and 2
(characterized by the index $\alpha_1$ and $\alpha_2$, respectively)
is given by
\be
\left(\frac{f_{12}}{f_1}\right)^{\alpha_1} = 
\left(\frac{f_{12}}{f_2}\right)^{\alpha_2}\,.
\label{Strans}
\ee
In general, high frequencies (say, above $\simeq 0.2 - 1$ Hz) are dominated 
by thermal noise, with the index $\alpha_3 = 2$; in the case of ULYSSES, 
this spectral region is also perturbed by the spacecraft rotational dynamics,
so that the data have been actually low-pass filtered to $5\times 10^{-2}\,{\rm Hz}
= f_{23}$ (for ULYSSES we can therefore assume $\alpha_3 \gg 1$ and we have formally achieved
that by setting $\alpha_3 = 10$).  
The frequency window between $f_{12}$ and $f_{23}$ can be characterized by a power law with 
index $\alpha_2  \approx - 0.5$, due to propagation noise \cite{pap0,pap1}. 
The analysis of past experiments has been 
mostly confined to this band, in particular to frequencies larger than $\approx 1/T = f_{12}$; 
in the case of ULYSSES, the procedure for the generation of residuals essentially cuts off 
all frequencies below  $\simeq 10^{-4}$ Hz; we have formally achieved 
this by adopting $\alpha_1 = -10$ (see Table \ref{tab;doppnoise}).

As we have already anticipated in the Introduction (and we shall show more in detail
in the next Sections), one of the main conclusions of this paper is that low frequencies 
(i.e. $\simlt 1/T$) are crucial to achieve large SNR and to widen the range of
accessible masses. The  noise spectrum for $f T\simlt 1$
is to a large extent unknown, as systematic effects, mostly 
related to the  orbit determination process and the tropospheric correction, 
are likely to come into play. For the goal of assessing the expected SNR for 
CASSINI we take the view that these systematic effects do not exceed the 
propagation noise above $ 10^{-4}$ Hz and keep $\alpha_2 = - 0.5$. 
The level $S_0$ is given by the main specification for the Allan deviation, i.e., 
$\sigma_y = 3\times 10^{-15} $ at $10^4$ sec, corresponding to 
$S_n(10^{-4} {\rm Hz}) = 9\times 10^{-26}$ sec; we also limit our band to a minimum frequency
of $10^{-6}\,{\rm Hz}$. For the interval up to $f_{12} = 10^{-4}\,{\rm Hz}$ the only experimental
evidences come from ULYSSES data, which showed an approximate behavior as $f^{-2}$
and, accordingly, we chose $\alpha_1 = - 2$; this value, however, is conservative, as in the
processing of ULYSSES data no particular attention was taken at the reduction of the noise
below $10^{-4}\,{\rm Hz}$. The
effect of possible different values of $\alpha_1$ will be explored in Sec. \ref{subsec;sensf}. 
Thermal noise, with $\alpha_3 = 2$, can be expected to set in at about $0.1\,{\rm Hz}$; 
however, the limited sampling time will prevent to go much beyond. 
In Fig. \ref{fig;doppnoise} we show the noise spectral density of CASSINI and ULYSSES,
computed using Eqs. (\ref{Sncol}) and (\ref{S0}) with the parameters given in
Table \ref{tab;doppnoise}.

Keeping $S_0$ and the frequencies $f_j$ (j=1,2,3) fixed, we now write the SNR 
in the form
\be 
\rho^2 = K\,\frac{\Mc^{5/3}T^{4/3}}{S_0 D^2}
\Upsilon (\xi_b, \xi_e, \theta; \alpha_j). 
\label{rho}
\ee 
The {\em sensitivity function} 
\ba
\Upsilon(\xi_b, \xi_e, \theta; \alpha_j)\equiv
K^{-1}\,\left\{\int_{\xi_b}^{\xi_e}\,
\xi^{-7/3}
\left[\sum_{j}\left(\frac{\xi}{\xi_j}\right)^{\alpha_j}\right]^{-1} \,
|{\tilde r}_{\theta}(\xi)|^2\,d\xi\right\}
\label{eff}
\ea
measures the change of SNR with respect to the "ideal" case, Eq. (\ref{rhoid}), and fulfils
\be
\Upsilon(0, \infty, \pi/2; 0) = 1\,.
\ee
Generally this function is less than unity and gives a degradation; we must however point out
that the value $\pi/2$ for the angle $\theta$ is not always (i.e. for any arbitrary choice 
of $\alpha_j$) the best; furthermore, we have defined the ``white noise" considering the level 
of the real noise at $f = 1/T$, so that $S_n(f)$ may actually be smaller than
the white noise for $f > 1/T$. Notice also that when $\alpha_1 \geq - 2/3$ the 
SNR spectral density is not integrable and the total SNR is dominated by the low frequency 
end $f_b$. 

The crucial physical parameter of a binary is its chirp mass $\Mc$, 
which characterizes the rate of change of the gravitational wave frequency and sets the 
relationship (except for class IV)
\ba
\Mc & =  & \frac{5^{3/5}}{(8\pi)^{8/5}}\, T_1^{-3/5}\,f_b^{-8/5} \,
\left [ 1 - \left (\frac{f_b}{f_e} \right )^{8/3} \right ]^{3/5} \nonumber\\
& = & \Mc_c \left [ 1 - \left (\frac{f_b}{f_e} \right )^{8/3} \right ]^{3/5}
\quad\quad ({\rm class\,\, I - III})\,.
\label{chirp1}
\ea
We see that, given $f_b$ and $T_1,$ there is a critical value
of the chirp mass
\ba
\Mc_c & = & \frac{5^{3/5}}{(8\pi)^{8/5}}\, T_1^{-3/5}\,f_b^{-8/5}
\nonumber\\
& \simeq & 9.2 \times 10^{5}\,
\left(\frac{f_b}{10^{-4}\,{\rm Hz}}\right)^{-8/5}\,
\left(\frac{T_1}{40\,{\rm days}}\right)^{-3/5}\,\Ms\,.
\label{chirpc}
\ea
above which no measurement is possible, in class I-to-III. At this critical value,
when $f_b \ll f_{\rm isco}$, $T_1$ basically coincides with the time to coalescence $\tau_c$.

When $\Mc \ll \Mc_c$, the signal is intrinsically narrow-band, and so are the measurements; 
the fractional bandwidth
\be
\Delta =\frac{f_e-f_b}{f_b}
\label{Delta}
\ee
is small and given by
\be
\Delta = \frac{3}{8}\,\left(\frac{\Mc}{\Mc_c}\right)^{5/3} = \frac{3}{8}\,\frac{T_1}{\tau_c}\,.
\label{Delta1}
\ee
This case, in which there is little power and information, is less interesting.

When the data set includes the merger (class IV), instead of Eq. (\ref{chirp1}) we have
\be 
\Mc = \Mc_c\, \left[\frac{T_1}{\tau_c} 
\left( 1 - \left (\frac{f_b}{f_{\rm isco}} \right )^{8/3}\right) \right]^{3/5}
\quad\quad ({\rm class\,\,\, IV})\,.
\label{chirpc1}
\ee
The coefficient within square brackets can be greater than unity, 
so that the inequality $\Mc < \Mc_c$ does not apply.

Since, with decreasing frequency, the spectral density of the SNR decreases   
or has a weaker growth below $1/T$, it is very convenient to have $Tf_b < 1$.
When $\xi_e = f_e T \ll 1$ the response  
$i\pi\xi\sin^2\theta$, Eq. (\ref{rfll1}), weakens the signal.  
When $\xi_e > 1$ the response introduces strong modulations that depend  
on $\theta$ and we have full sensitivity.  
This is the interesting region and will be explored in Sec. \ref{subsec;sensf} using the  
dynamical behavior of the source.  

Before discussing in detail the behavior of the SNR in observations of in-spiral
signals as function of the relevant parameters, it is instructive to compare 
the sensitivity of Doppler experiments to the radiation emitted during the in-spiral and 
merger phase. If $\rho_{\rm insp}$ and $\rho_{\rm merg}$ correspond to 
the optimal SNR achievable by observations of the same source in the two regimes,
from Eq. (\ref{rhoE}) we can write:
\be
\frac{\rho_{\rm insp}}{\rho_{\rm merg}} = 
\left[\frac{
\int_{f_{\rm B}}^{f_{\rm isco}}\,
\frac{\overline{|\rf_{\theta}(f)|^2}}{f^2\,S_n(f)}\,
\frac{dE_{\rm insp}(f)}{df}\,df}
{\int_{f_{\rm isco}}^{f_{\rm qnm}}\,
\frac{\overline{|\rf_{\theta}(f)|^2}}{f^2\,S_n(f)}\,
\frac{dE_{\rm merg}(f)}{df}\,df}
\right]^{1/2}\,.
\label{rhoim}
\ee
We stress that in the previous relationship we have assumed optimal 
filtering not only for the in-spiral but also for the merger signal; if for the former 
the assumption is realistic, cfr. \cite{Sathya}, for the latter it is indeed
very optimistic. Notice also that $\rho_{\rm insp}$ is the largest SNR
obtainable for a chirp, as we integrate over the widest frequency
band $(f_B, f_{\rm isco})$ for a data set of length $T_1$ (cfr. Sec. \ref{subsec;sensf}), 
and we have averaged the three-pulse response function over the angle 
$\theta$, cfr. Eq (\ref{rfaver}).
For a given source, if $(\rho_{\rm insp}/\rho_{\rm merg}) > 1$ ($< 1$)
it is therefore 'more convenient' to search for it
by means of observations of the in-spiral (merger) signal. 

\begin{figure}
\centerline{\mbox{\epsfysize=15.cm
\epsffile{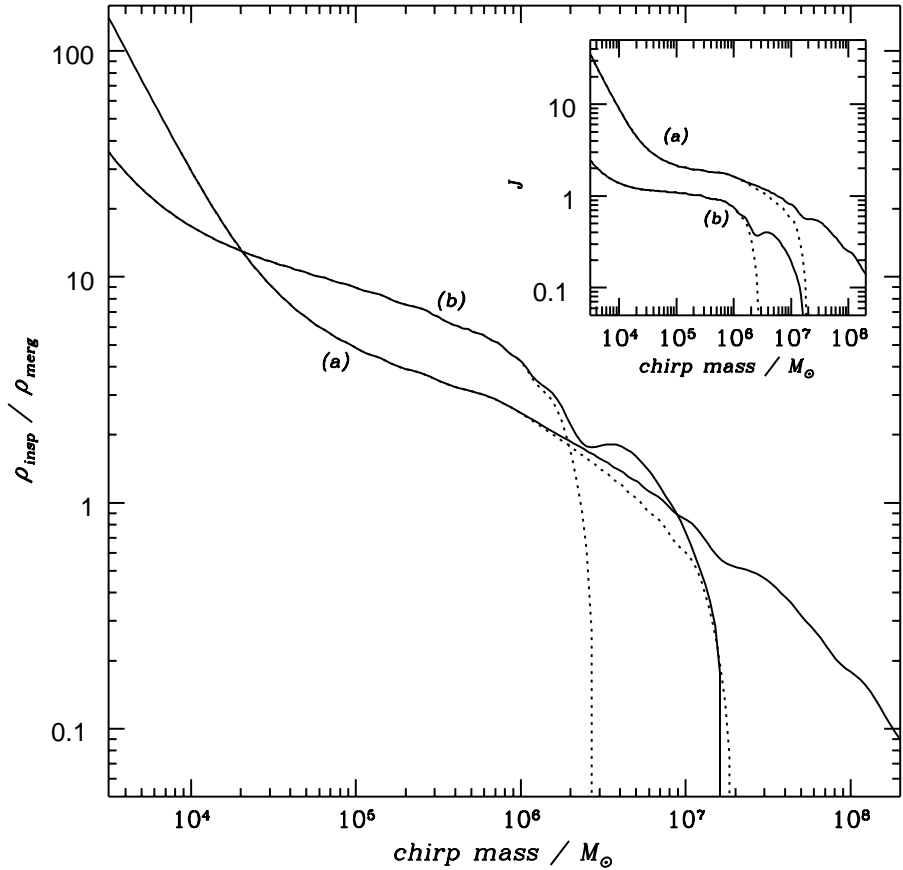}}}
\caption{\label{fig;rhoim}
Comparison of the sensitivity of CASSINI experiments
in observations of in-spiral and merger signals. For a given source,
the plot shows the ratio of the optimal SNR achievable in observations
of the in-spiral signal, $\rho_{\rm insp}$, with
respect to that produced during the merger phase, $\rho_{\rm merg}$, as a function of the 
chirp mass and for the mass ratios 1 and 0.01, label
(a) and (b) respectively. The solid lines refer to the low frequency cut-off 
$10^{-6}$ Hz, whereas the dotted lines to $10^{-4}$ Hz. The small box shows,
for the same cases, the relative overlap $J$, Eq. (\ref{Upsim}), of the in-spiral and merger 
signal with respect to the instrument response. The time of observation is assumed to
be $T_1 = $ 40 days and the noise spectral density 
$S_n(f)$ is computed according to Eqs. (\ref{Sncol}) -- (\ref{Strans}) and 
Table \ref{tab;doppnoise}.
}
\end{figure}

Inserting Eqs. (\ref{dEmdf}) and (\ref{dEinspdf}) into Eq. (\ref{rhoim}), the ratio
of SNR's becomes
\be
\frac{\rho_{\rm insp}}{\rho_{\rm merg}} \simeq 1.1 \,J\,
\left(\frac{\epsilon_{\rm merg}}{0.1}\right)^{-1/2}\,
\left(\frac{T}{10^4\,{\rm sec}}\right)^{1/6}\,
\left(\frac{M}{2\times 10^7\,\Ms}\right)^{1/3}\,
\left(\frac{\mu}{5\times 10^6\,\Ms}\right)^{-1/2}\,,
\label{rhoim1}
\ee
where
\be
J \equiv
\left[\frac{
\int_{(T\,f_{\rm B})}^{(T\,f_{\rm isco})}\,
\frac{\overline{|\rf_{\theta}(\xi)|^2}}{\xi^{7/3}\,S_n(\xi)}\,d\xi}
{\int_{(T\,f_{\rm isco})}^{(T\,f_{\rm qnm})}\,
\frac{\overline{|\rf_{\theta}(\xi)|^2}}{\xi^2\,S_n(\xi)}\,d\xi}
\right]^{1/2}
\label{Upsim}
\ee
measures the relative overlap of the in-spiral and merger signals with respect
to the instrument response. In Fig. \ref{fig;rhoim} we show 
$(\rho_{\rm insp}/\rho_{\rm merg})$ and $J$
as a function of the chirp mass for two different mass ratios, 1 and
0.01, for a nominal CASSINI experiment of 40 days with $S_n(f)$ given by
Eqs. (\ref{Sncol} -- \ref{Snorm}). The value of the chirp mass $\Mc \sim
10^7\,\Ms$ (for a low frequency cut-off of $10^{-6}$ Hz) sets the transition between the systems 
that are more convenient to be searched during the in-spiral phase and 
those during the merger phase: for $\Mc > 10^7\,\Ms$, $\rho_{\rm merg}$
rapidly exceeds $\rho_{\rm insp}$ by a factor $\sim 10$ or more. Assuming a low frequency 
cut-off around $10^{-4}$ Hz, as done in past experiments, the transition occurs at a lower
value of $\Mc \simeq 5\times 10^{6}\,\Ms$ and the drops of 
$(\rho_{\rm insp}/\rho_{\rm merg})$ is much steeper. As a consequence,
for $\Mc \simlt 10^7\,\Ms$ the neglect of the merger signal in searching
templates is not serious; for $\Mc \simgt 10^7\,\Ms$ coherent search 
techniques that include merger waveforms would rapidly increase the instrument range of sight, by more than one order
of magnitude. However, at present, the radiation produced during the final burst does not seem to be
the most promising signal to be searched for in Doppler experiments,
as astrophysical considerations suggest that its amplitude is not
large enough to dominate the noise and matched filtering techniques
would fail to enhance the SNR due to the poor prior knowledge of the
waveform. Only coincidence experiments can offer a chance of carrying out
such searches; this work is currently in progress -- 
under the assumption of very simple shapes of the waves -- 
for the analysis of data recorded in coincidence during the 
tracking of ULYSSES, MARS-OBSERVER and GALILEO \cite{UMOGcoinc}.

\subsection{Sensitivity function} 
\label{subsec;sensf}

In this Section we discuss the dependence of the sensitivity function 
$\Upsilon(\xi_b, \xi_e, \theta; \alpha_j)$ on the main parameters
that characterize the in-spiral signal and the instrument, in particular the conditions 
under which $\Upsilon$ attains a value close to the ideal one or, equivalently, the 
possible causes of degradation.

\begin{figure}
\centerline{\mbox{\epsfysize=15.cm
\epsffile{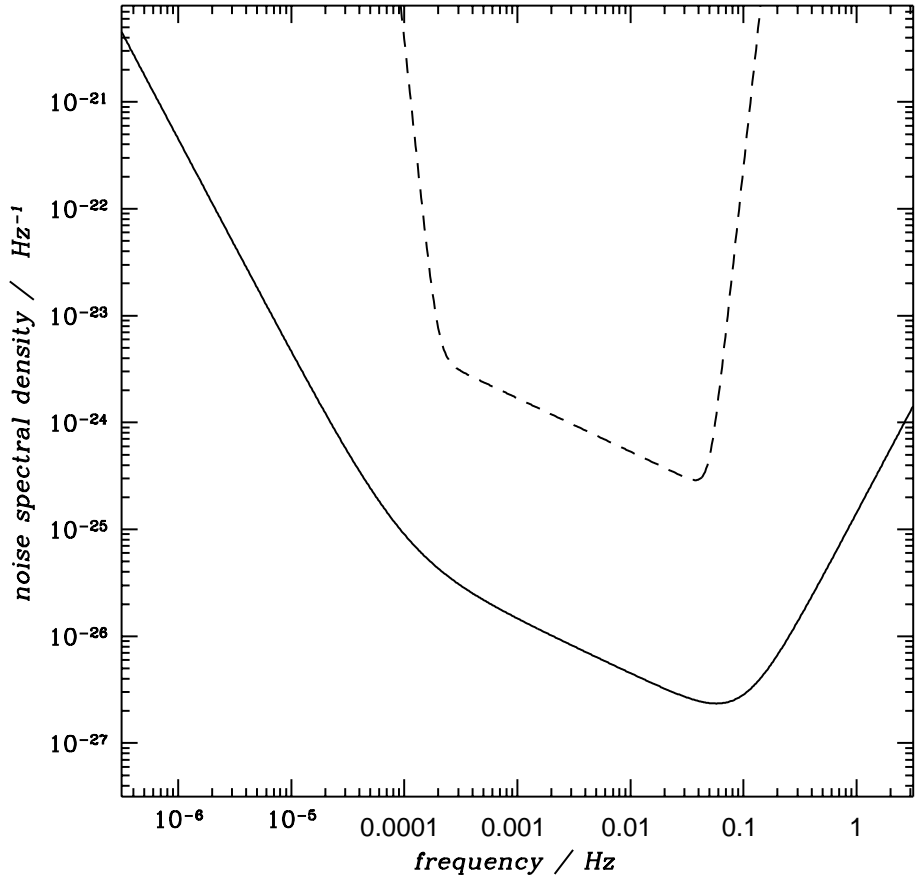}}}
\caption{\label{fig;doppnoise}
Noise spectral density of CASSINI (solid line) and ULYSSES
(dash-line). $S_n(f)$ is computed according to Eqs. (\ref{Sncol}) -- (\ref{Strans}) and 
Table \ref{tab;doppnoise}.
}
\end{figure}

\begin{figure}
\centerline{\mbox{\epsfysize=15.cm
\epsffile{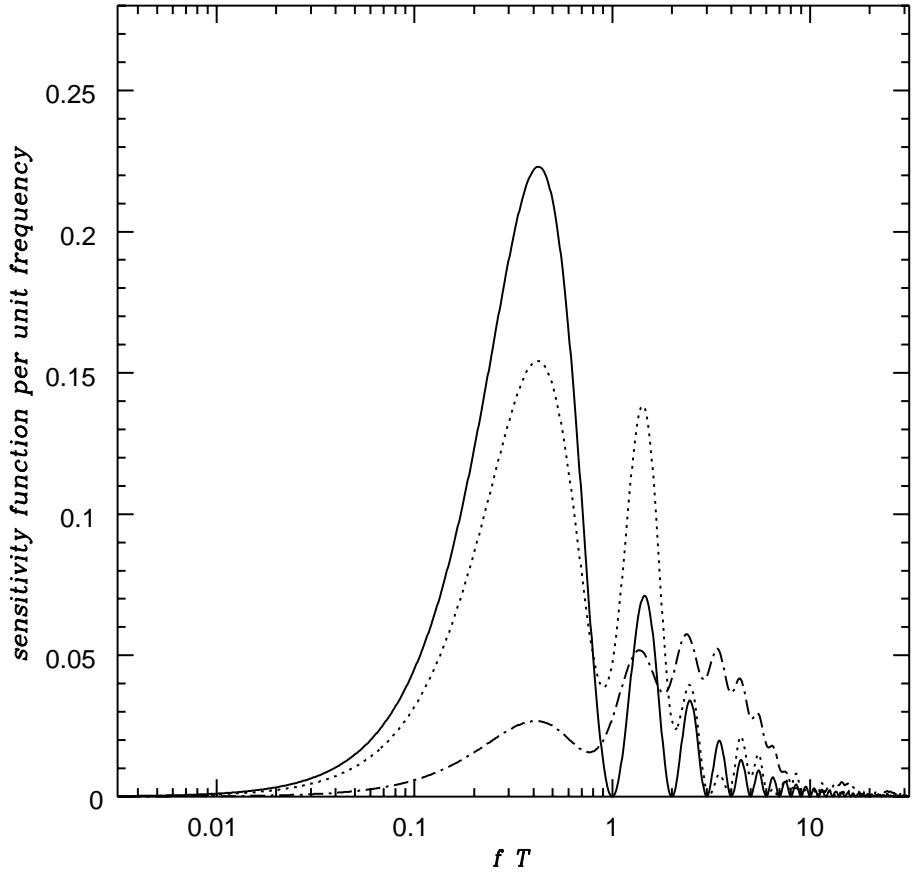}}}
\caption{\label{fig;dupsdx}
The spectrum $\tilde\Upsilon (f)$ of the sensitivity function, see Eq. (\ref{eff}). 
We have used CASSINI's noise parameters, Table \ref{tab;doppnoise} and Fig. 
\ref{fig;doppnoise} (solid line: $\cos\theta = 0$;
dotted line: $\cos\theta = 0.4$; dotted-dashed line: $\cos\theta = 0.8$).
}
\end{figure}

\begin{figure}
\centerline{\mbox{\epsfysize=15.cm
\epsffile{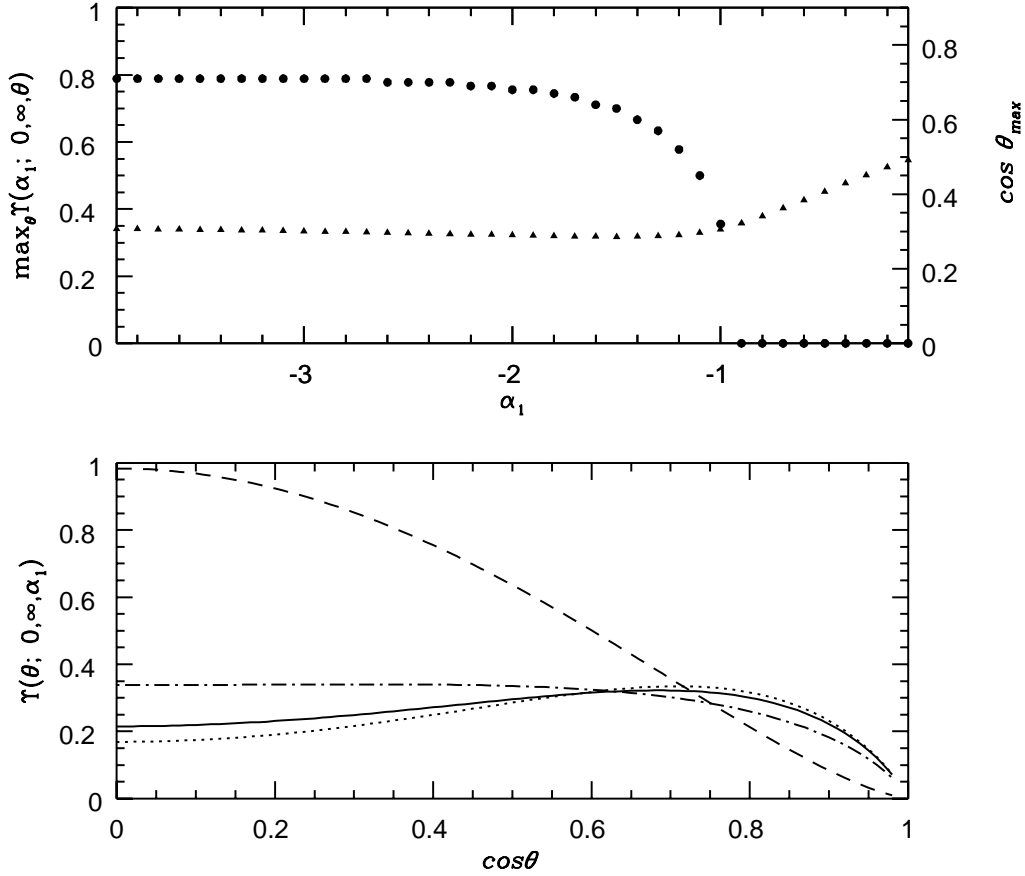}}}
\caption{\label{fig;peak}
The deterioration of the sensitivity function $\Upsilon$, and therefore of the SNR,
due to the angle $\theta$ between the source and the probe and the redness of the spectrum
at low frequencies,
given by $\alpha_1$ (cfr. Eqs. (\ref{Sncol}) -- (\ref{Strans}) and Table
\ref{tab;doppnoise}), 
as described by the function $\Upsilon(\xi_e = 0, \xi_b = \infty, \theta; \alpha_1)$.
In the upper panel we show, as a function of $\alpha_1$, the maximum value of 
$\Upsilon(\xi_e = 0, \xi_b = \infty, \theta; \alpha_1)$ (triangles) and the angle
$\theta_{\rm max}$ (bullets) at which it is attained. The `normal' value  
$\theta_{\rm max}= \pi/2$ occurs above $\alpha_1 \simeq -0.9$.
The lower panel shows the behavior of $\Upsilon(\xi_e = 0, \xi_b = \infty, \theta; \alpha_1)$
as a function of $\cos\theta$ for the case $\alpha_1 = 0$ 
(dashed line), $\alpha_1 = -1$ (dotted-dashed line), $\alpha_1 = -2$
(continuous line) and $\alpha_1 = -3$ (dotted line); the values of $\alpha_2$ and
$\alpha_3$ are $-1/2$ and $+2$, respectively, corresponding to those of 
the CASSINI experiments. Note that, for $\alpha_1 \le -1$,
the smallest deterioration does not occur at 
the obvious value $\theta = \pi/2$; in particular, for $\alpha_1 = -2$ it occurs
at $\cos\theta \simeq 0.68$, so that a small 
signal enhancement with respect to the ideal case is possible.
}
\end{figure}

To illustrate the first reason of degradation -- the unfavorable angle $\theta$ -- as well
as the behavior of $\Upsilon$ depending on the frequency range swept by the signal during
the observation time, 
we show in Fig. \ref{fig;dupsdx} the sensitivity function per unit frequency 
${\tilde \Upsilon}(f)$ for CASSINI's nominal noise (see Fig. \ref{fig;doppnoise} and
Table \ref{tab;doppnoise}). The sensitivity peak lies  
in the interval $0.1 \simlt \xi \simlt 1$, and broadens and diminishes as $\cos\theta$ 
increases. For $\xi\ll 1$, ${\tilde \Upsilon}(f)$ is a monotonic, decreasing function   
of $\cos\theta$, while as $\xi$ exceeds unity, oscillations of   
increasing frequency occur. The overall effect of the angle $\theta$ on $\Upsilon$ 
is given by the function $\Upsilon(\xi_e = 0, \xi_b = \infty, \theta; \alpha_1)$, 
plotted in Fig. \ref{fig;peak}, where we also keep $\alpha_1$ as free parameter in
order to show the effect of the redness of the noise spectral density at low
frequencies (that is $f \simlt 1/T$);
for $\alpha_1 \simgt -0.9$, $\Upsilon(\xi_e = 0, \xi_b = \infty, \theta; \alpha_1)$ 
has a maximum at the expected value $\theta = \pi/2$. This is not true anymore for
redder spectra, and the value of $\theta$ at which $\Upsilon$ reaches the maximum
depends on $\alpha_1$; for $\alpha_1 = -2$ it corresponds to $\cos\theta \simeq 0.68$.

\begin{figure}
\centerline{\mbox{\epsfysize=15.cm
\epsffile{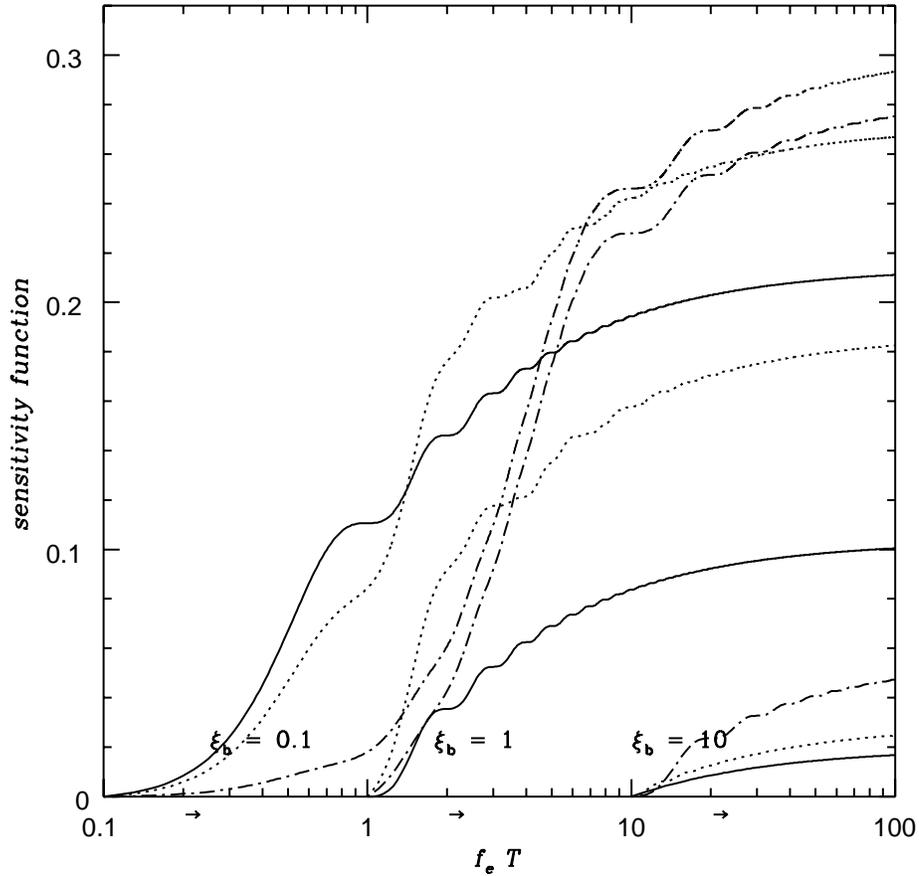}}}
\caption{\label{fig;effxi}
The behavior of the sensitivity function $\Upsilon$ as a function of the 
final frequency $f_e$ (in units of the round-trip light-time $T$) 
for selected values of the initial 
frequency $T\,f_b = \xi_b = 0.1\,,1\,,10$ (see labels).
CASSINI's noise parameters have been used (Table \ref{tab;doppnoise} and Fig. \ref{fig;doppnoise}).
To the right of the arrows we have broad band searches with $f_e - f_b > f_b$ (solid
line: $\cos \theta = 0$; dotted line: $\cos \theta = 0.4$;
dotted-dashed line: $\cos \theta = 0.8$).
}
\end{figure}

Concerning the second reason of degradation of $\Upsilon$ 
-- the limited bandwidth of the signal --
we show first (Fig. \ref{fig;effxi}) the effect of $f_b$ and 
$f_e$. The deterioration can be strong: $\Upsilon$ can significantly 
deviate (by one order of magnitude or 
more) from the ideal value. When both $\xi_b$ and $\xi_e$ are $\simlt 1$, $\Upsilon$ is   
a smooth function of $f_b$ and $f_e$, but as soon as they enter the region  
$f \simgt 1/T$, there are oscillations. We have used the `nominal' CASSINI form of
the noise spectrum (Fig. \ref{fig;doppnoise}) and we keep it fix from now on.
  
It is interesting to investigate $\Upsilon$ in relation to relevant 
astrophysical parameters. We concentrate on four fiducial "best" sources defined by
choosing the initial frequency $f_b$ and assuming that the coalescence occurs at the
end of the record, so that $f_b = f_B$ and $f_e = f_{\rm isco}$. Each of them is a binary with equal masses
($\eta = 1/4$), which determines the chirp mass
\be  
\Mc = \Mc_c \left[1-\left(\frac{f_b}{f_{\rm isco}}\right)^{8/3}\right]^ {3/5},  
\ee 
in terms of the critical mass for a nominal run of 40 days, see Eq. (\ref{chirpc}).
This sets also the final frequency $f_e = f_{\rm isco}$ and the number of wave
cycles recorded at the detector. Table \ref{tab;systems} summaries the main properties of
the fiducial sources. In all four cases $f_e\gg f_b$, so that we have a wide 
band search and the source's 
chirp mass is only slightly smaller than the critical value. 

\begin{figure}
\centerline{\mbox{\epsfysize=15.cm
\epsffile{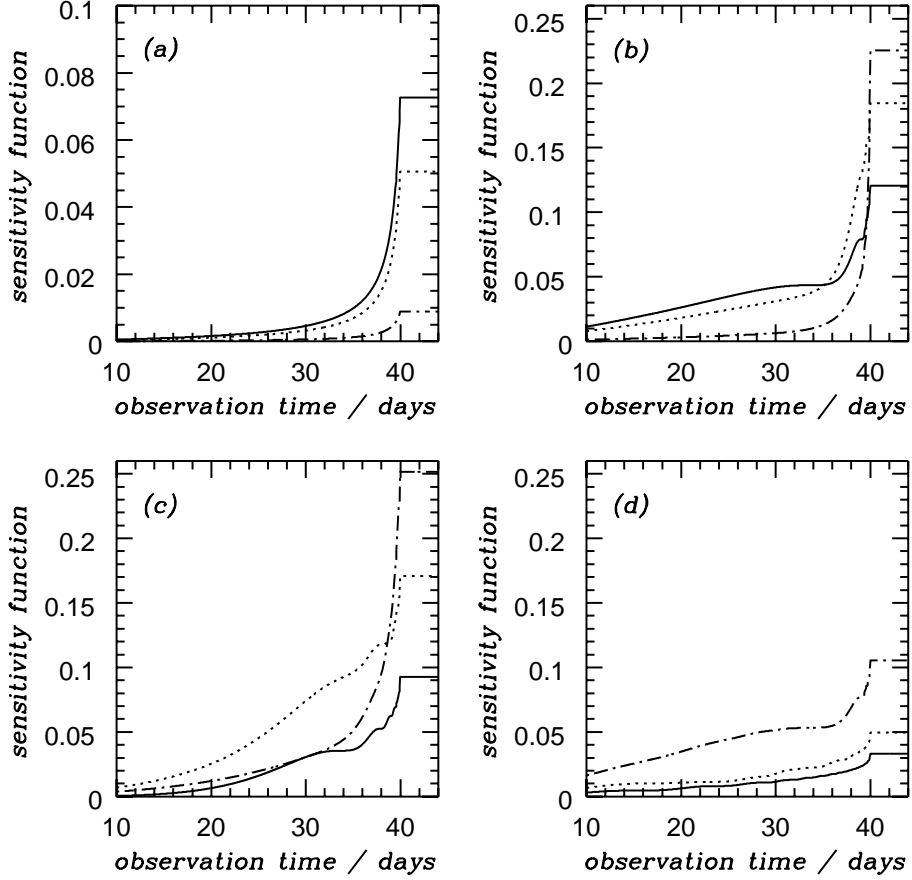}}}
\caption{\label{fig;effT1}
The behavior of the sensitivity function $\Upsilon$ as a function of $T_1$, 
for the four fiducial sources described in the text and in Table \ref{tab;systems} 
(panels (a) to (d),
respectively), with $f_b = 10^{-5}\,,5\times 10^{-5}\,,10^{-4}\,,5\times 10^{-4}\,{\rm Hz}$,
equal masses, coalescence at the end of the record and chirp mass almost equal to its
the critical value for a 40 days run. The deterioration as $T_1$ decreases reflects
the loss of signal-to-noise ratio due to the narrowing of the band-width (solid 
line: $\cos \theta = 0$; dotted line: $\cos \theta = 0.4$;  
dotted-dashed line: $\cos \theta = 0.8$). Notice the different scale in panel (a).
}
\end{figure}

Keeping $f_b$ and $\Mc$ fixed, we first study the effect of $T_1$ on $\Upsilon$.
As $T_1$ decreases, the signal becomes of class III 
and we loose a progressively greater part of its end. Insisting that the chirp mass remains 
the same, the final frequency $f_e$ decreases according to Eq. (\ref{fe}). This 
progressive restriction of the search bandwidth is responsible for the great 
deterioration in $\Upsilon$, and therefore in SNR, shown in Fig. \ref{fig;effT1}.
Since we have assumed that there is no signal beyond the coalescence frequency 
${f_{\rm isco}}$, the sensitivity function remains constant for $T_1 > 40$ days.  

\begin{figure}
\centerline{\mbox{\epsfysize=15.cm
\epsffile{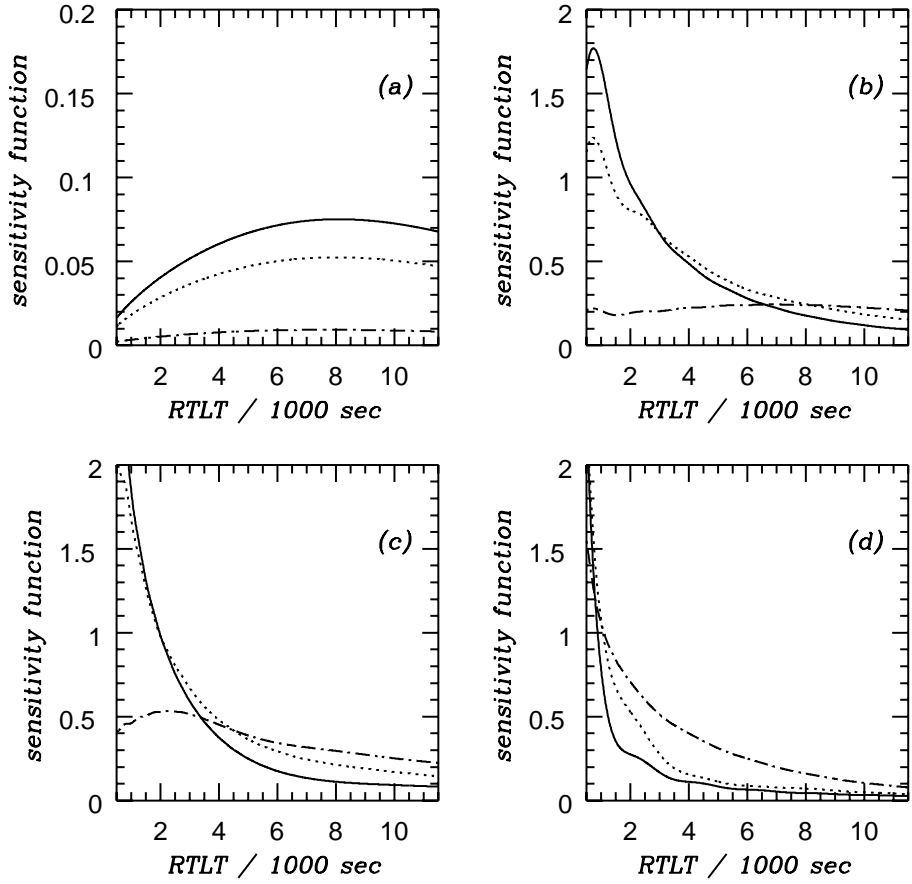}}}
\caption{\label{fig;effrtlt}
The behavior of the sensitivity function $\Upsilon$ as a function of
the round-trip light time for same four fiducial sources of the previous Figure
(solid 
line: $\cos \theta = 0$; dotted line: $\cos \theta = 0.4$;  
dotted-dashed line: $\cos \theta = 0.8$). Notice the different scale in panel (a).
}
\end{figure}

In Fig. \ref{fig;effrtlt} we show how the sensitivity function 
is affected by the round-trip light-time 
$T$, for the same fiducial sources (and with a run of 40 days) . 
The variation of $T$ deplaces the three-pulse response function -- Eq. (\ref{rf2}) and
Fig. \ref{fig;dupsdx} --
in relation to the search band $(f_b,f_e)$. The deterioration with increasing $T$ is 
due to the fact $\xi_e$ 
becomes smaller than unity and one looses signal from the high frequency side.

\begin{figure}
\centerline{\mbox{\epsfysize=15.cm
\epsffile{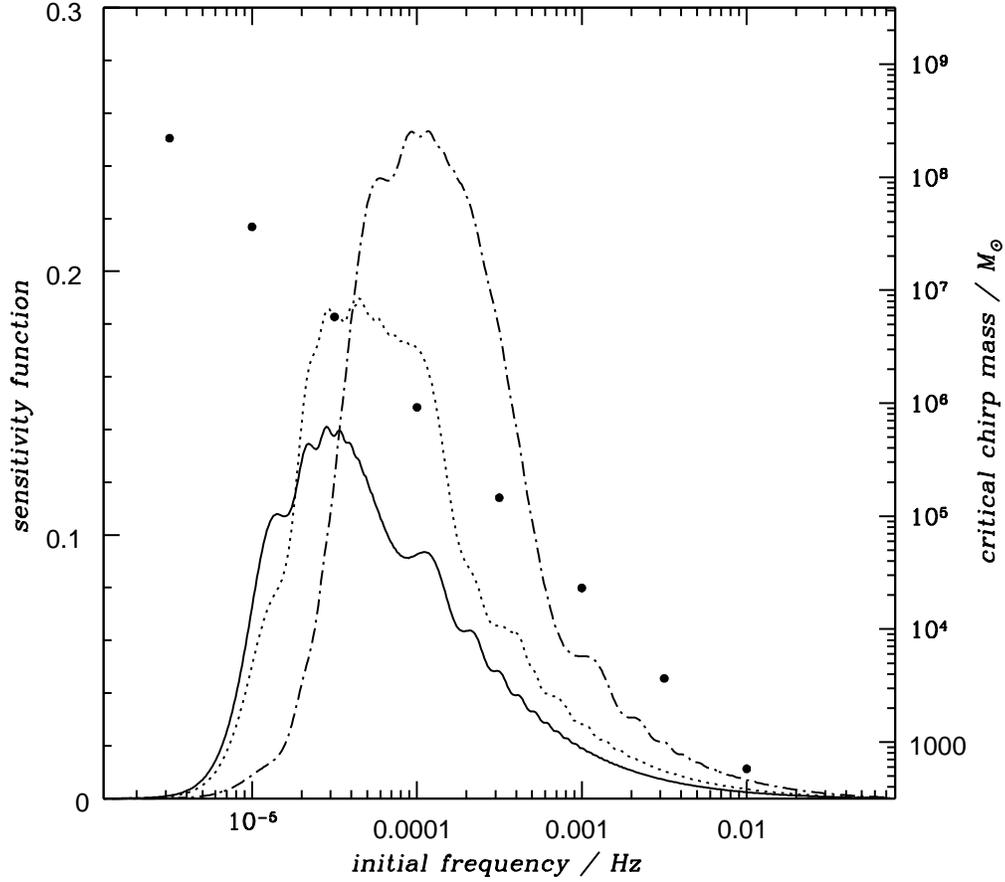}}}
\caption{\label{fig;effchirp}
The behavior of the sensitivity function as a function of the initial frequency
$f_b$ for wide band searches of class III, with $f_b = f_B$, $f_e = f_{\rm isco}$
and $T_1 = 40\,{\rm days}$. 
The corresponding value of the chirp mass -- near to
the critical value (\ref{chirpc}) -- is  given by the filled circles
(solid line:
$\cos\theta = 0$; dotted line: $\cos\theta = 0.4$;
dotted-dashed line: $\cos\theta = 0.8$).
}
\end{figure}

Finally, in Fig. \ref{fig;effchirp} we investigate the effect of $f_b$ (that is the epoch 
of the binary life at which the instrument is "turned on") on $\Upsilon$; for each value
of $f_b$, we have in fact computed the corresponding critical chirp mass (\ref{chirpc})
for a 40-day experiment and then evaluated the sensitivity function for three different
source's orientation in the sky; essentially, for each $f_b$ we tune the observation on
the mass that produces the highest SNR. The largest values of $\Upsilon$ are achieved for 
$10^{-5}\,{\rm Hz} \simlt f_b \simlt 10^{-3}\,{\rm Hz}$, corresponding to 
$10^4\,\Ms\simlt \Mc_c \simlt 10^8\,\Ms$.

\subsection{Sensitivity of past and future Doppler experiment}  
\label{subsec;senscass}
 
In this Section we analyze the sensitivity of Doppler experiments -- in particular CASSINI --
to search for in-spiral signals; we estimate  
the range of sensitivity and the achievable signal-to-noise ratio as function of the chirp mass $\Mc$ and
the initial frequency $f_b$ (as well as $\eta$).  
We consider in detail two targets, the Virgo Cluster ($D \simeq 17\,{\rm Mpc}$)
and the Galactic Centre ($D \simeq 8$ kpc); the results are presented so that 
they can be scaled to an 
arbitrary distance and location in the sky.
  
\begin{figure}
\centerline{\mbox{\epsfysize=15.cm
\epsffile{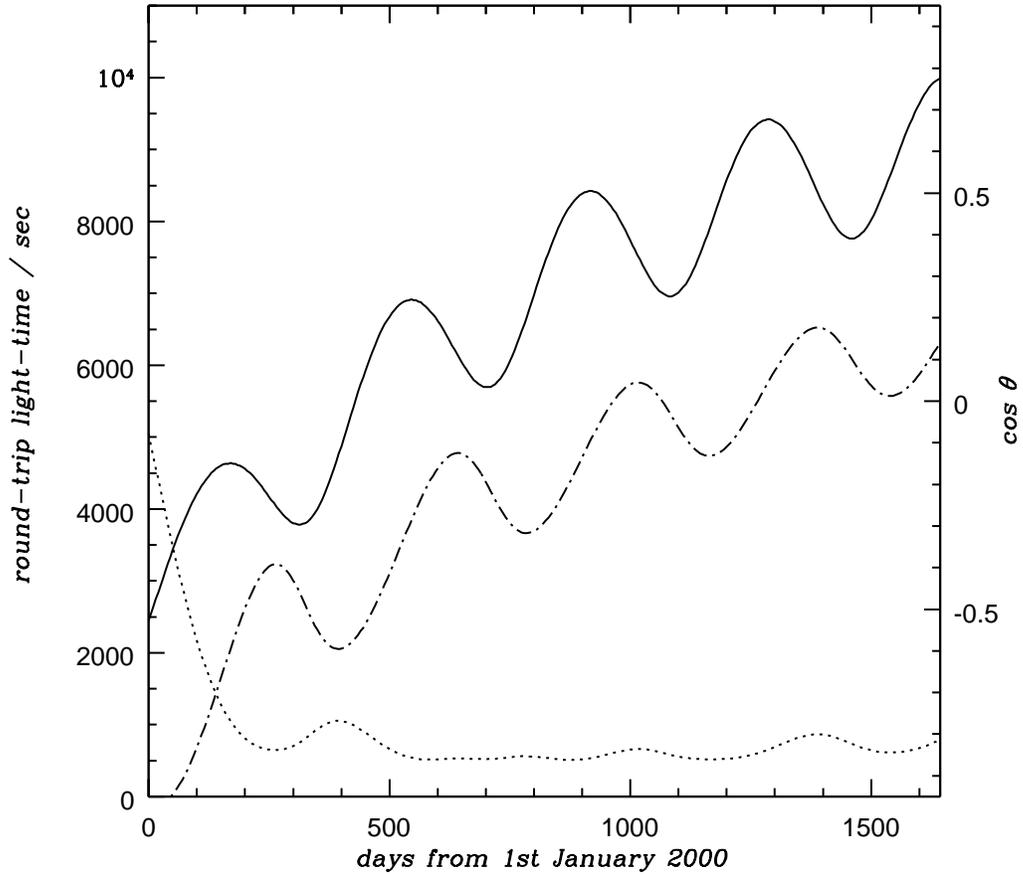}}}
\caption{\label{fig;casspar}
CASSINI's cruise to Saturn. The solid line gives the round-trip light-time $T$ as function
of time and we show also the variation of $\cos\theta$ for the Virgo Cluster 
(dashed-dotted line) and the Galactic Centre (dotted line). The three gravitational wave 
experiments take place during the days 695-735, 1071-1111 and 1443-1483 from 
1st January 2000.
}
\end{figure}

CASSINI will perform three experiments, each lasting 40 days at the epoch of  
three solar oppositions, i.e. when the Sun, the Earth and the probe  
are approximately aligned, in this order, so as to minimize the noise  
due to interplanetary plasma (see Table \ref{tab;doppexp} and Fig. \ref{fig;casspar}) \cite{cas}.  
The main tracking will be done in X band from the operational antennas  
of NASA's Deep Space Network (DSN), and in K$_a$ band  
from the new experimental DSN 
station DSS25 at Goldstone (California).  Another advanced station to be build 
in Sardinia (the Sardinia Radio Telescope) has been recently funded and it
is hoped that it will track CASSINI in K$_a$-band (up and down) and X-band
(down).
The radio system of CASSINI's mission has a specification for the overall  
Allan deviation in K$_a$ band of $3\times 10^{-15}$, for integration  
times between 1000 and 10,000 sec, one-to-two orders of magnitude better  
than previous experiments. In addition, 
the sensitivity will be improved, with respect to past missions, 
by the longer time of observation and, possibly, by the combination of the  
three records. Fig. \ref{fig;casspar} shows the round-trip light-time and
the value of $\cos\theta$ for the Virgo Cluster 
and the Galactic Centre during the cruise of the space-probe to Saturn.

\begin{figure}
\centerline{\mbox{\epsfysize=15.cm
\epsffile{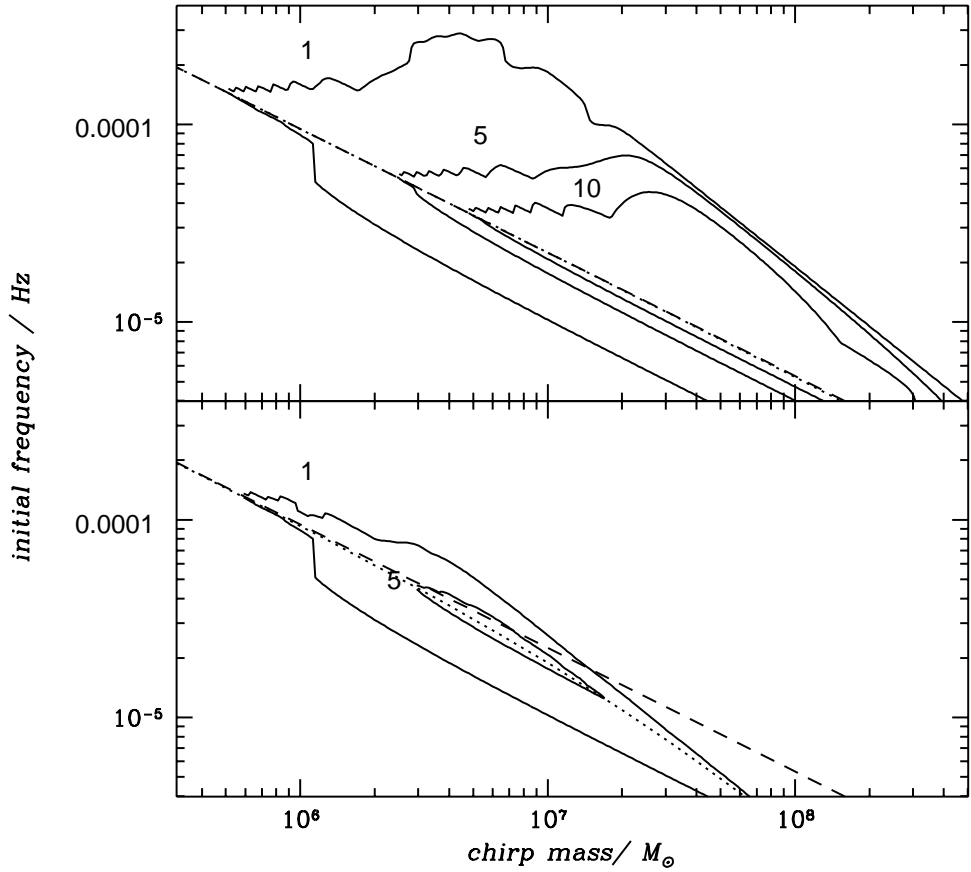}}}
\caption{\label{fig;virgocass40}
Detectable in-spiral signals emitted by 
binary systems in the Virgo Cluster for the third CASSINI's experiment of 40 days.    
The contour plots of the 
SNR (in amplitude) for the levels 1, 5 and 10 are shown. The top and bottom panels refer to 
the mass ratios $M_2/M_1 = 1$ and 0.01, respectively. In the latter case no signals are observable at $\rho \ge 10$. 
The dashed diagonal line gives the critical chirp mass $\Mc_c$, see Eq. (\ref{chirpc}).
The dotted line marks the transition between signals of class III and IV. Notice that 
for $M_1 = M_2$ (upper panel) the two lines almost coincide.
}
\end{figure}

We begin by investigating the expected SNR produced by in-spiraling signals
of possible sources located in the Virgo Cluster in the plane $(\Mc,f_b)$.  
Since it turns out that the three planned experiments have comparable  
sensitivities, we give explicit results only for the last one, in 2004. 
Fig. \ref{fig;virgocass40} shows for the Virgo Cluster three different levels of SNR:  
$\rho = 1\,,5\,{\rm and}\,10$.  
It confirms the theoretical suggestion that the detectable events of  
class III lie in a narrow strip below the critical value of the chirp  
mass. All events above it are of class IV and include the merger. This plot shows  
how the SNR changes as function of both $\Mc$ and $f_b$: at $\rho = 1$ and for black holes  
of comparable masses, the smallest detectable chirp mass is  
$\simeq 5\times 10^5 \; \Ms$. Lower masses are excluded because the systems are
too far from coalescence and masses larger than $\sim 10^9\,\Ms$ are not
accessible because their emission  
frequencies are too low. Note the drastic curtailment if a low frequency  
cut off at $1/T \sim 10^{-4}$ Hz is assumed, as it was done in the past with  
ULYSSES; the strongest signals fall below $f_bT = 1$.  
The SNR is very much reduced when a small mass ratio is assumed  
(lower panel); an important and unknown factor in assessing the detectability 
of gravitational radiation from MBHB's is the statistics of this ratio.  
 
One wonders if a source in the Virgo cluster detected in the last opposition  
is traceable also in the second or the first (albeit at a smaller SNR),  
thus adding important phase information. We checked that for  
$T_1 = 40\,{\rm days}$ this is not the case: detectable sources are too  
near coalescence, in relation to the (roughly) one year separation between different  
runs.  
We have then tried doubling the observation time to $T_1 = 80$ days  
(`extended' experiments) and interestingly found that for some  
sources the three extended data sets can be combined: the strongest sources 
detectable in one opposition are also visible in the previous one.
A single, extended experiment does not show drastic improvements over the  
nominal record; however, the accessible portion of the plane $(\Mc,f_b)$ becomes slightly wider.
Of course,
a drastic improvement in sensitivity and detection probability
would be expected for a continuous run of a few years.
 
\begin{figure}
\centerline{\mbox{\epsfysize=15.cm
\epsffile{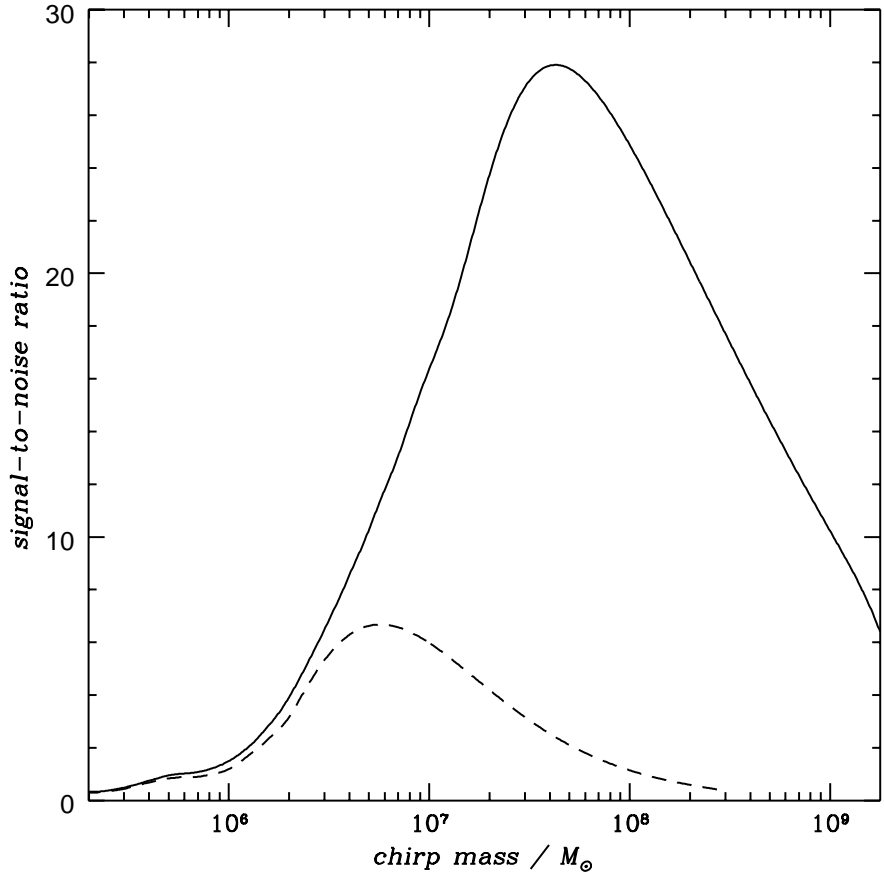}}}
\caption{\label{fig;virgocassmax}
Signal-to-noise ratio for  in-spiral signals emitted by binaries in the Virgo cluster (D = 17 Mpc)
with coalescence at the end of the record -- that is  $f_b = f_B$ and $f_e = f_{\rm isco}$ --
as a function of the chirp mass. The solid line refers to 
the case $M_1 = M_2$, while the dashed line
to $M_2/M_1 = 0.01$. We have considered CASSINI third opposition and $T_1 = 40\,{\rm days}$.
}
\end{figure}

How far is the farthest detectable system? From Fig. \ref{fig;virgocassmax} we see that the highest SNR,  
about 28, produced by a source in the Virgo cluster is attained for a system of comparable masses  
$\simeq 5\times 10^7 \, \Ms$; at $\rho = 1$ this corresponds  
to a distance $\approx 600 $ Mpc; if $M_2/M_1 = 10^{-2}$, the ken is reduced by a factor
$\simeq 5$. 

\begin{figure}
\centerline{\mbox{\epsfysize=15.cm
\epsffile{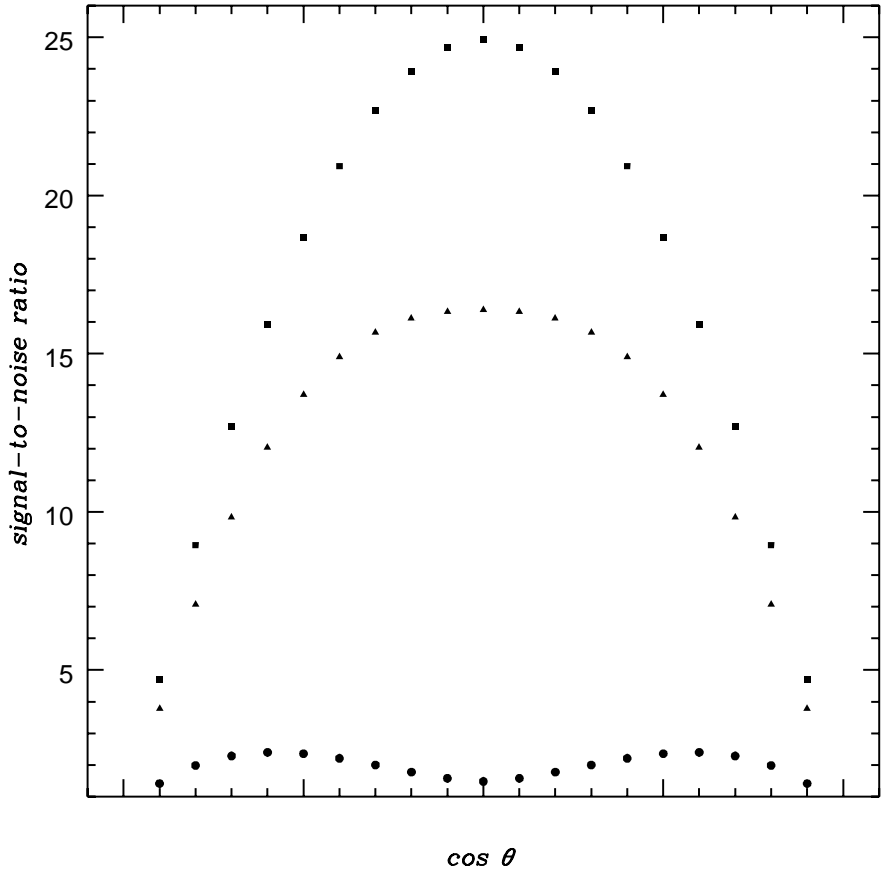}}}
\caption{\label{fig;allsky}
The signal-to-noise ratio as a function
of $\cos\theta$ during CASSINI third opposition, $T_1 = 40\,{\rm days}$, 
for in-spiral signals emitted by binary systems located at the same distance
of the  Virgo cluster ($D = 17\,{\rm Mpc}$) with $f_b = f_B$ and
$f_e = f_{\rm isco}$, that is final coalescence at the end of the record
(squares: $\Mc = 10^8\,\Ms$; triangles: $\Ms = 10^7\,\Ms$;
bullets: $\Ms = 10^6\,\Ms$; for all $\eta = 1/4$). The decrease for $|\cos\theta|\rightarrow 1$
is of course due to the approach to the case of propagation along the instrument's
beam; the saddle in correspondence of $|\cos\theta| \simeq 0$
for $\Mc = 10^6\,\Ms$ is due to the fact that
the frequency band swept by the signal encounters a minimum
of the response function at $f T \approx 1$, which vanishes for
$\cos\theta = 0$, see Figs. \ref{fig;doppresp} and \ref{fig;dupsdx}.
}
\end{figure}

In order to have an idea of the change in SNR  
due to different angles $\theta$, in Fig. \ref{fig;allsky} we have studied this  
dependence for the third experiment in three different cases: $\Mc = 10^6\,,10^7$
and $10^8\,\Ms$ ($f_b = f_B$, $f_e = f_{\rm isco}$ and $\eta = 1/4$). Note that,
for  $\Mc = 10^6\,\Ms$, $\cos\theta = 0$ is a local minimum, rather than 
the expected maximum: this is due to the fact that signals fall in one of the
high frequency minima of the three-pulse response, see Figs. \ref{fig;doppresp} and \ref{fig;dupsdx}.
Figs. \ref{fig;virgocass40}, \ref{fig;virgocassmax} and \ref{fig;allsky}
therefore allow 
to have an overall picture of the all-sky sensitivity of CASSINI, see Eqs. (\ref{rho}) and 
(\ref{eff}).

\begin{figure}
\centerline{\mbox{\epsfysize=15.cm
\epsffile{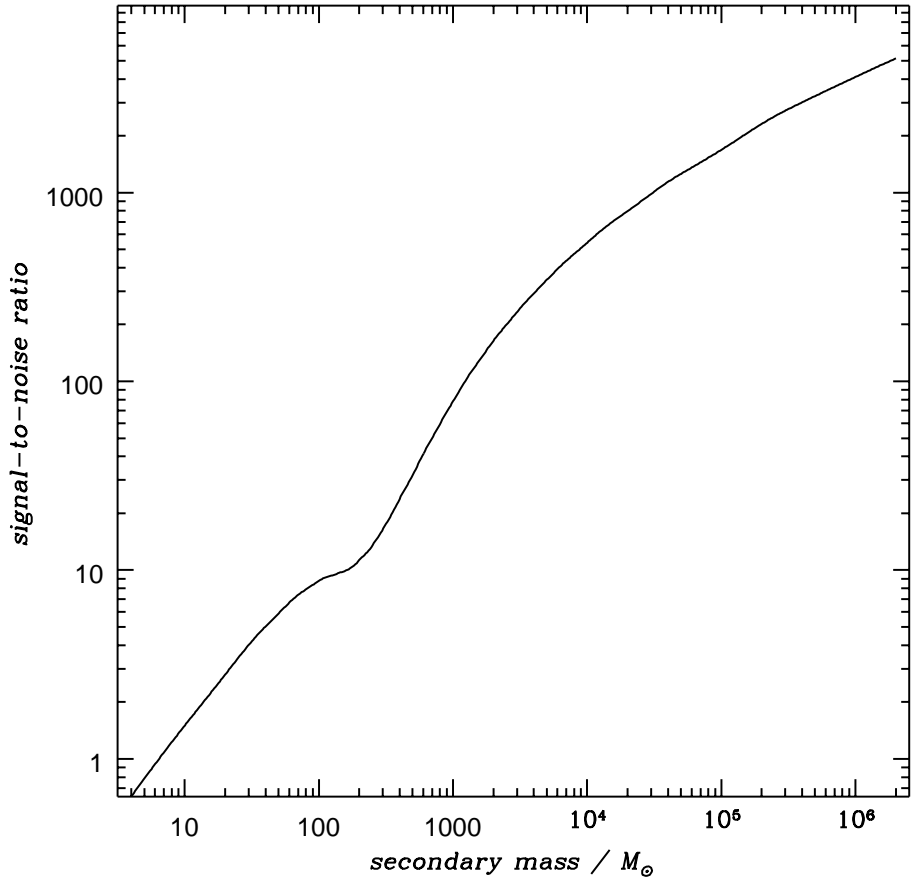}}}
\caption{\label{fig;gccassmax}
Expected optimal signal-to-noise ratio produced in 
CASSINI records by in-spiraling binaries in the Galactic Centre.
The plot shows the maximum SNR -- {\it i.e.} computed for $f_b = f_B$ and 
coalescence
at the end of the record ($T_1 = 40$ days), therefore $f_e = f_{\rm isco}$ -- 
produced by binary systems with $M_1 = 2\times 10^6\,\Ms$ and 
a secondary black hole of mass $M_2\le M_1$ 
as a function of $M_2$, during CASSINI's third opposition.
}
\end{figure}

We turn now to galactic observations and consider a binary system in the  
centre of Our Galaxy composed of a primary black hole of   
mass $M_1 = 2\times 10^6\,\Ms$, whose existence is strongly inferred from observations
carried out over the past 25 years \cite{EG96}, and a secondary, smaller object of mass $M_2$. 
In Fig. \ref{fig;gccass40} we show for the third CASSINI opposition
the SNR contour levels $\rho = 1\,,5\,,100$ in the plane $(M_2,f_b)$.  Of course,
there is no signal above 
$f_{\rm isco} \simeq 2 \times 10^{-3} \,(M_1/2\times 10^6 \,\Ms)^{-1}\,{\rm Hz}$, cfr. Eq. 
(\ref{fisco}), essentially independent of $M_2$. CASSINI proofs to be a fairly sensitive
instruments for such signals; in fact, each of the three experiments is sensitive
to secondary objects of mass down to $\sim 10-50\,\Ms$ and for $M_2 \simgt 100\,\Ms$ the expected
SNR can be $\gg 10$, see Fig. \ref{fig;gccassmax}. Remarkably, most of the sources 
visible during one experiment are  detectable also in the others; in fact the time to 
coalescence is $\simeq   2995\,(f_b/5\times 10^{-4}\,{\rm Hz})^{-8/3}\,(100\,\Ms/\mu)\,  
(2\times 10^6\,\Ms/M)^{2/3}\,{\rm days}$ (cfr. also Fig. \ref{fig;detprob}). If a candidate
signal in spotted in one of the data sets, it is therefore possible to chase it also in the
other oppositions, which provides a robust method of accepting or discarding candidate
events by coherently tracking the gravitational phase over a period of some years.
This is a direct consequence of the fact that many signals would be recorded as
class II and III (therefore far from the final coalescence) as is shown in 
Fig. \ref{fig;gccass40} by the dotted and dashed line in the plane $(M_2,f_b)$; this is
in contrast with what appends for observations in the Virgo cluster, where most
of the signals show up as class IV. 
The good sensitivity for nearby sources allows also to search for possible binary systems
in other galaxies of the Local Group.

It is interesting to compare CASSINI's sensitivity with past   
Doppler experiments (Table \ref{tab;doppexp}). The best Doppler data available today come from  
observations carried out in 1991, 1992 and 1993 with the probes ULYSSES,  
GALILEO and MARS-OBSERVER. In 1993 all three spacecraft have been  
simultaneously tracked for about two weeks,  
with a triple coincidence experiment whose data are being analyzed;  
this experiment will be useful to search for signals coming from the merger phase \cite{UMOGcoinc}.  
 
We concentrate on ULYSSES' 1992 experiment \cite{pap0,pap1}.
After pre-processing, only 14 days of data have been retained, centered around the sun 
opposition.  Their Allan deviation (at 1000 sec) is variable, with only its best  
value near the specification, $3 \times 10^{-14}$. Data outside the  
bandwidth $\simeq 2.3\times 10^{-4}\, - \,5\times 10^{-2}\,{\rm Hz}$ have been  
discarded, see Fig. \ref{fig;doppnoise}: at high frequency they are corrupted by 
the anomalous rotation of the spacecraft; at low frequency,
the subtraction of orbital and media contributions  
was carried out with a fit, which made the data unreliable below $f \simeq 1/T$. 
In the intermediate band the noise spectrum is well  
approximated by a power-low $\propto f^{-0.5}$ (see Table \ref{tab;doppnoise}). 
The low frequency cut off limits the largest observable mass to 
\be
\Mc \simlt 9\times 10^{6}\,\left(4\,\eta\right)^{3/5}\,\Ms\,,
\label{chirpmax_u}
\ee
and the value of the critical chirp mass (\ref{chirpc})
for the 14-days ULYSSES second opposition experiment is
\be
\Mc \simeq 1.3\times 10^{5}
\left(\frac{f_b}{5\times 10^{-4}\,{\rm Hz}}\right)^{-8/5}\,
\left(\frac{T_1}{14\,{\rm days}}\right)^{-3/5}\,\Ms\,;
\label{chirpc_u}
\ee
Eqs. (\ref{chirpmax_u}) and (\ref{chirpc_u}) clearly indicate that 
the most interesting region in the ($\Mc, f_b$) plane
is not accessible (cfr. Fig. \ref{fig;virgocass40}) and therefore the largest 
attainable distance is strongly reduced. Moreover, the noise spectral density, as defined
by Eqs. (\ref{Sncol})--(\ref{Strans}) with the parameters reported in Table
\ref{tab;doppnoise} (see Fig. \ref{fig;doppnoise}), is about two orders
of magnitude larger than for CASSINI. According to Eq. (\ref{ken}),
the ideal {\em ken} (we recall that it is defined for $\rho_{\rm id} =1$) holds
\be 
D_{\rm id} \simeq 1\,\left(\frac{\Mc}{1.3\times 10^5 \Ms}\right)^{5/6}\,
\left(\frac{S_0}{2.5\times 10^{-24}\,{\rm Hz}^{-1}}\right)^{-1/2}
\,\left(\frac{T}{4430\,{\rm sec}}\right)^{2/3}\,{\rm Mpc}\,.
\ee 
Therefore, in conventional astrophysical scenarios, ULYSSES' experiments are relevant 
only for the Galactic Centre, whose angular 
position with respect to the spacecraft was $\theta \approx 109^o$. 
It is straightforward to check that at $\rho = 5$
 secondary black holes with $M_2 \simgt 10^3\,\Ms$ would be detectable.

\section{CONCLUSIONS}  
 
We have analyzed the sensitivity of Doppler detectors to gravitational waves   
emitted by coalescing binaries in their in-spiral phase. For  CASSINI's experiments   
sources in the Virgo Cluster would be observable at a signal-to-noise ratio 
up to  $\simeq 30$.   
Furthermore, binary systems with  $M_1 \sim M_2 \sim 5\times 10^7\,\Ms$ at distance
$\sim 600\,{\rm Mpc}$ are 
within the range of the instrument. If in the centre of Our Galaxy there is a massive   
black hole of  $2\times 10^6\,\Ms$, orbiting black holes with mass $\simgt 50\,\Ms$ 
would be easily detectable; for 
larger masses the signal-to-noise ratio can be considerably higher and therefore
one can reach other galaxies within the Local Group. 
These are considerable   
improvements in sensitivity and range of accessible masses with respect to past Doppler experiments.  
  
This analysis has been made in relation to astrophysically reasonable sources,   
but without any concern for the probability of such events and their detection.  
Considering our ignorance about sources of gravitational waves, in particular for   
events in galactic nuclei, this point of view appears reasonable, especially for   
non dedicated experiments like CASSINI's; however, it is healthy and sobering to   
assess what one can predict on the basis of {\em  very simple and conventional   
astrophysical models}.  
The estimation of the event rate is 
a much debated problem in low frequency gravitational wave   
experiments (see \cite{LISA,SR,Hae,Lisavecch,BH}   
and reference therein), with sensitivity not good enough to detect signals from   
known stellar mass galactic binaries. We briefly consider here two models: (a) accretion   
of small black holes onto a massive one in a galactic core and (b)   
coalescence of two MBHB's of comparable mass, possibly resulting from the   
mergers of two galaxies. In the context of Doppler experiments, they are appropriate 
to small and large distances, respectively.  
  
\begin{figure}
\centerline{\mbox{\epsfysize=15.cm
\epsffile{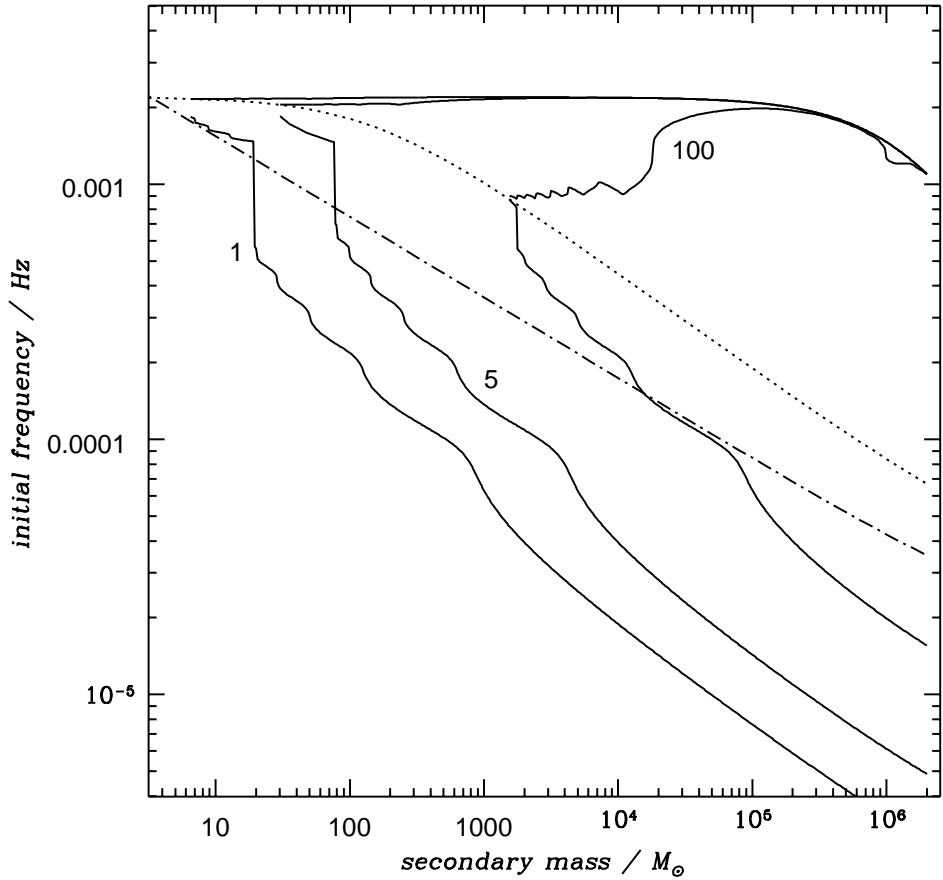}}}
\caption{\label{fig;gccass40}
Detectability of radiation emitted by the in-spiral of a secondary black hole of 
mass $M_2$ orbiting a primary object with
mass $M_1 = 2\times 10^6\,\Ms$, in the Galactic   
Centre (D $\simeq$ 8 Kpc) during CASSINI's third experiment ($T_1 = 40\,{\rm days}$). 
The contour plots of SNR $ = 1\,,5$ and 100 in the plane $(M_2,f_b)$ are shown. 
The dashed-dotted line divides the plane $(M_2,f_b)$ in signal of class II (below) 
and
III (above), while above the dotted line the signals are registered as of 
class IV.
}
\end{figure}

\begin{figure}
\centerline{\mbox{\epsfysize=15.cm
\epsffile{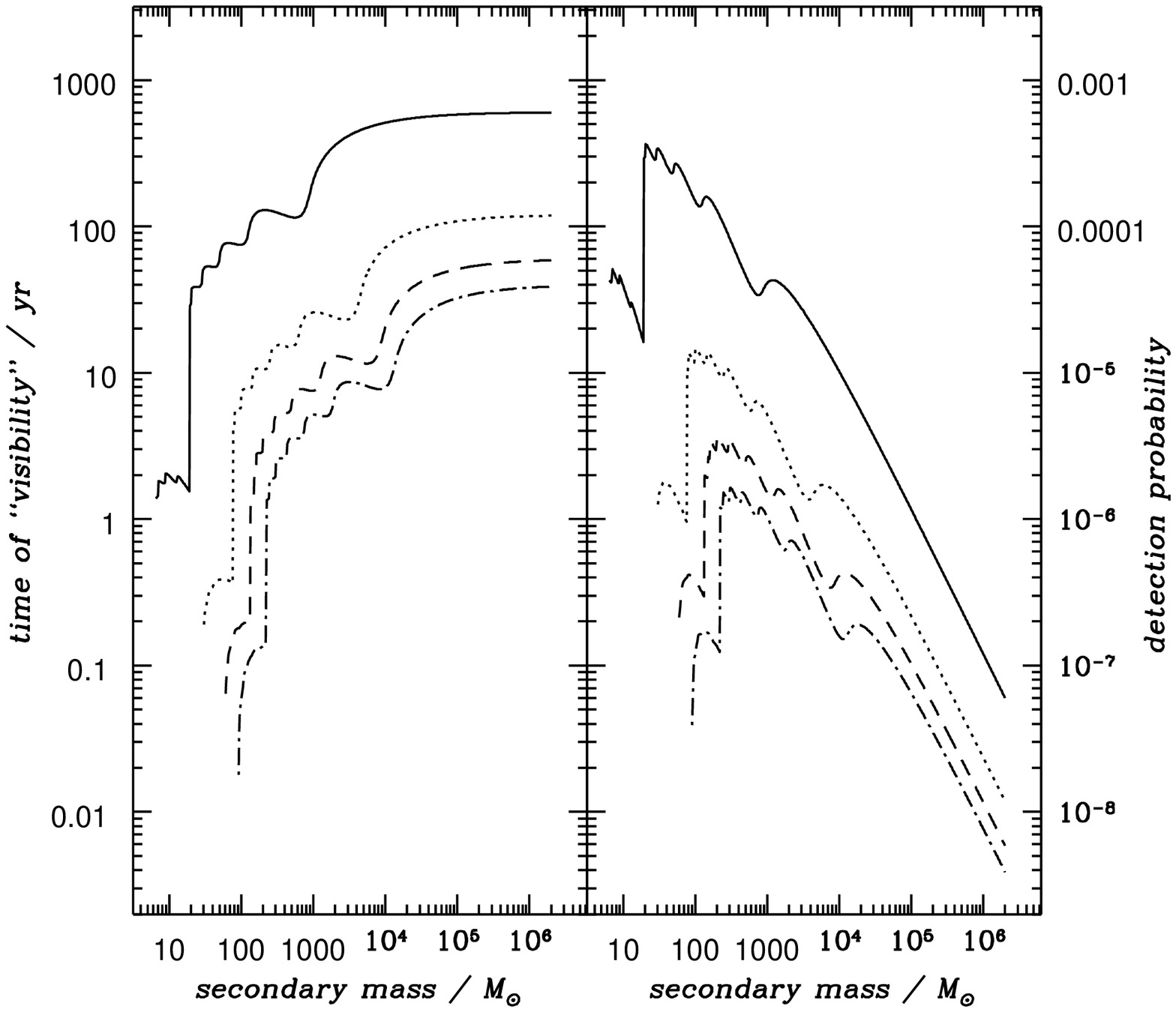}}}
\caption{\label{fig;detprob}
Time of `visibility' $T_D$ and probability of success for the detection
of radiation emitted by the in-spiral of a secondary black hole into
a primary of mass $M_1 = 2\times 10^6\,\Ms$ in Our Galactic Centre during
CASSINI third opposition. The curves
refer to four different thresholds of detection ($\rho_{\rm D} = 1$:
solid line; $\rho_{\rm D} = 5$: dotted line; $\rho_{\rm D} = 10$: dashed line;
$\rho_{\rm D} = 15$: dotted-dashed line).
}
\end{figure}

Let consider now case (a).
Suppose that a central black hole $M$ in a typical galaxy (i.e., in Our Galactic Centre) 
has captured $N= M/M_2$ masses $M_2$ at the uniform rate $R = H_0 M/M_2$ during the whole 
life of the Universe $H_0^{-1} \simeq 10^{10}\,h_{100}\,{\rm yr}$, where $h_{100}$ is the Hubble
constant normalized to $100\,{\rm km}\,{\rm sec}^{-1}\,{\rm Mpc}^{-1}$; each of these captures gives rise 
to a chirp. In this ideal case, in which the spread in mass is neglected, one could consider 
the search as follows. For a given $M_2$ (and fixed $M_1$, which determine uniquely the
chirp mass) there is a characteristic initial frequency $f_B$, given by Eq. (\ref{fbmax}),
for which we have a wide band detection and coalescence at the 
end of the record; this corresponds to the situation in which the SNR is maximum. 
As $f_b$ diminishes, the fractional bandwidth $\Delta$ -- see Eq. (\ref{Delta}) -- decreases
according to
\be 
f_b \simeq f_B [1 - (1- \Delta)^{-8/3}]^{3/8}\,,\nonumber
\ee
and so does the SNR, proportionally to $\Delta$:
\be
\rho^2 \simeq \rho^2_{\rm id} \,K^{-1}\,{\tilde \Upsilon}(f)\, f \,\Delta\,,\nonumber
\ee
where $f\sim f_b\sim f_e$.
Since a small value of $\Delta$ increases the useful time for detection $T_{\rm D}$ --
that is the total time for which the source is ``visible'' above the detection
threshold $\rho_{\rm D}$ -- detection will generally occur at a SNR less than the maximum 
(attained for $f_b = f_B$ and $f_e = f_{\rm isco}$)
and with a start frequency $f_b < f_B$. $\rho_{\rm D}$ determines the threshold 
$\Delta_{\rm D}$ for 
the fractional bandwidth and $T_{\rm D}$, via Eq. (\ref{Delta1}). Indeed, 
the success probability is $P = R\,T_{\rm D}$ and we have computed it using the
results reported in Fig. \ref{fig;gccass40}. Its behavior as a function of the mass $M_2$, 
which enters through both $R$ and $T_{\rm D}$, 
is given in Fig. \ref{fig;detprob} for the thresholds $\rho_{\rm D} = 1$,
5, 10 and 15. This allows us to determine
the most favorable value of $M_2$, about $100\,\Ms$, and 
the corresponding (optimistic!) success probability, in each experiment of $T_1 = 40$ days,
is $\sim 10^{-5}$.

We turn now to case (b). To go to very large distances and to reach the 
largest number of sources, we must aim at large masses and use low frequencies, in the 
band $f\,T\simlt 1$. In this frequency range, the noise spectral density is well described
by a single power low 
\be
S_n(f) \simeq S_0\,\left(\frac{f}{f_1}\right)^{\alpha_1}\,.
\ee
The SNR produced by a source that sweeps the bandwidth $\Delta$, where
$\xi_e = (1 + \Delta)\,\xi_b$, see Eq. (\ref{Delta}), can be analytically computed
from Eq. (\ref{rho}) assuming the simple low-frequency behavior of the 
Doppler response filter  (\ref{rf2ll1}):
\be
\rho^2 = k\,\frac{T^{4/3}\,\Mc^{5/3}}{S_0 \,D^2}\,\left(T\,f_1\right)^{\alpha_1}\,
(T\,f_b)^{(2/3 - \alpha_1)}\,\left[\left(1 + \Delta\right)^{(2/3 - \alpha_1)} - 1\right]\,;
\label{rhoprob}
\ee
$k$ is a numerical coefficients that depends on the angles $\iota\,,\varphi$
and $\theta$. The largest attainable distance, in observations of a binary of mass $\Mc$
is therefore 
\be
D = k^{1/2}\,\frac{T\,\Mc^{5/6}}{S_0^{1/2} \,\rho}\,f_1^{\alpha_1/2}\,
f_b^{(1/3 - \alpha_1/2)}\,\left[\left(1 + \Delta\right)^{(2/3 - \alpha_1)} - 1\right]^{1/2}
\ee
and the corresponding number of detectable events
\ba
N & \sim & R\,T_1\,(D\,H_0)^3 \nonumber\\
& = & R\,T_1\,H_0^3\,k^{3/2}\,\frac{T^3\,\Mc^{5/2}}{S_0^{3/2} \,\rho^3}\,f_1^{3\,\alpha_1/2}\,
f_b^{1 - 3\,\alpha_1/2}\,\left[\left(1 + \Delta\right)^{2/3 - \alpha_1} - 1\right]^{3/2}\,.
\label{Nevents}
\ea
Of course, the range of the instrument depends also on the mass of the
source and one can express $\Mc$ as a function of $f_b$, $T_1$ and $\Delta$,
using Eq. (\ref{chirp1}); Eq. (\ref{Nevents}) therefore becomes:
\ba
N & = & R\,H_0^3\,k'\,\frac{T^3\,\Mc^{5/2}}{S_0^{3/2}\,T_1^{1/2}\,\rho^3}\,f_1^{(3\,\alpha_1/2)}\,
f_b^{-3\,(1 + \alpha_1/2)}\nonumber\\
& & \times \left\{
\left[\left(1 + \Delta\right)^{(2/3 - \alpha_1)} - 1\right]
\left[1 - \left(1 + \Delta\right)^{-8/3} \right]
\right\}^{3/2}\,,
\ea
where $k'$ is a numerical constant depending on $k$. We see that $N$ 
is proportional to $f_b^{-3 (1 + \alpha_1/2)}$ and increases
with decreasing frequencies if $\alpha_1 < -2$. The nominal case $\alpha_1 = -2$ (Table II)
marks the transition between the two alternatives, whether it is useful or not to go
to lower frequencies. If we assume (naively and optimistically) that each galaxy in the 
Universe contains a massive black hole and has undergone
$N_{\rm m}$ merging events leading to the formation of a binary with time to
coalescence shorter than the
Hubble time, we can estimate the event rate as $R = N_{\rm m} N_{\rm g}/H_0$, where $N_{\rm g}$
is the total number of galaxies. As we have seen in the previous section, the {\it ken}
of CASSINI is $D \sim 600\,{\rm Mpc}$ and therefore the probability of detection is
\be
P \sim 10^{-1}\,N_{\rm m}\,\left(\frac{3\,T_1}{120\,{\rm days}}\right)\,
\left(\frac{D}{600\,{\rm Mpc}}\right)^3\,\left(\frac{H_0^{-1}}{3\,{\rm Gpc}}\right)^{-3}\,.
\ee
Of  course, larger distances could be attained if we knew the waveform produced during
the final collapse and the merger could be included in the search.   
In the present situation of ignorance, however, a good level of confidence in the 
detection of merger signals requires coincidence experiments.  

We conclude that, in conventional astrophysical scenarios, the events we  
are looking for do not have a large probability to be detected, the  
extragalactic case (b) being more favorable. Note, however, that for low
chirp mass sources $\Mc \sim 100\,\Ms$, the total SNR could be increased
by combining two or three records.

Present Doppler experiments are relatively cheap, but use non dedicated  
spacecraft and are limited in  
instrumentation, orbits, observation time and by several  
other constraints. However, despite their low sensitivity, they are the only one that at present
can access the coalescence of binaries involving massive objects, and therefore 
quite valuable. They might discover gravitational waves;
if they do not, they will provide interesting astrophysical limits on massive sources,
in particular in Our Galactic Centre, the Local 
Group and the Virgo Cluster. Since real data are already available 
in large amounts and the
search techniques are very similar to those
implemented for Earth-based (and, in future, space-borne) laser interferometers,
Doppler experiments can also be used to test filtering and processing techniques.

\acknowledgments

We are very grateful to J. W. Armstrong for his continuous collaboration in Doppler experiments  
and his comments. We also thank H. D. Wahlquist and F. B. Estabrook for extensive discussions.
A.V. acknowledges a {\it Carlo A. Sacchi} fellowship and the kind hospitality of 
the University of Wales College of Cardiff during the early stage of this work.

%
%

\begin{table}[c]
\caption{\label{tab;doppexp}
Summary of Doppler experiments to search for
gravitational waves. $T_1$ is the effective length of data available;
$\sigma_y$ the average effective Allan variance of the data
and $T$ the average round-trip light-time of the radio link during 
the data acquisition 
runs. The frequency bands of the 
up-link (S: $2.1\,{\rm GHz}$, X: $7.2\,{\rm GHz}$, K$_a$: $34\,{\rm GHz}$)
and down-link (S: $2.3\,{\rm GHz}$, X: $8.4\,{\rm GHz}$, K$_a$: $32\,{\rm GHz}$)
carriers are shown in the two final columns.
For the four  short 
experiments carried out with PIONEER 10 and 11 and VOYAGER 1 see [21]
and references therein.
}
\begin{tabular}{ l c | c c c  c  c}
           &  &                   &                      &                      &        &           \\
 Space-probe          &  &   $T_1/{\rm days}$  & $\sigma_y$  &  $T/{\rm sec}$ & up-link & down-link \\
           &   &                    &                      &                      &        &           \\
\hline
           &   &                    &                      &                      &        &           \\
ULYSSES 3-Dec-90/4-Jan-91 &   &   3.5              &      $3\times 10^{-14}$               &    600               &   S    &   S,X     \\
ULYSSES 20-Feb-92/18-March-92 &   &   14          &      $7\times 10^{-14}$                &    4430              &   S    &   S,X     \\
ULYSSES Mar-93/Apr-93 &  &    19         &      $1.4\times 10^{-13}$          &    4100              &   S    &   S,X     \\
           &    &                   &                      &                      &        &           \\
\hline
           &   &                    &                      &                      &        &           \\
GALILEO Mar-93/Apr-93 &   &   19         &      $2.3\times 10^{-13}$              &    945               &   S    &   S       \\
           &   &                    &                      &                      &        &           \\
\hline
           &  &                     &                      &                      &        &           \\
MARS-OBSERVER Mar-93/Apr-93  &   &   19          &      $5\times 10^{-14}$               &    1150              &   X    &     X     \\
           &    &                   &                      &                      &        &           \\
\hline
           &   &                    &                      &                      &        &           \\
CASSINI 26-Nov-01/5-Jan-02 &  &     40              &      $3\times 10^{-15}$              &    5783              &  X,  K$_a$ & X,  K$_a$ \\
CASSINI 7-Dec-02/16-Jan-03 &   &    40              &      $3\times 10^{-15}$             &    7036              &  X,  K$_a$ & X,  K$_a$ \\
CASSINI 15-Dec-03/24-Jan-04 &   &    40              &     $3 \times 10^{-15}$            &    7859              &  X,  K$_a$ & X,  K$_a$
\end{tabular}
\end{table}

\begin{table}[c]
\caption{\label{tab;doppnoise}
Characteristic parameters of the ULYSSES and CASSINI noise spectral density. 
The table shows the reference frequency
$1/T$; the reference noise spectral density level $S_0 = S_n(1/T)$; the transition frequencies
$f_{12}, f_{23}$; the exponents $\alpha_j$ and the frequencies $f_j$
($j = 1,2,3$) of each regime, see Eqs. (\ref{Sncol}) -- (\ref{Snorm}). 
When, as in our case, $T\,f_{12} = 1$, $T\,f_{23} \gg 1$ and $\alpha_3 > \alpha_2$ then
$Tf_1 = 2^{1/\alpha_1}$, 
$Tf_2 = 2^{1/\alpha_2}$ and
$Tf_3 = Tf_{23}\, 2^{1/\alpha_3}\,(T f_{23})^{-\alpha_2/\alpha_3}$.
}
\begin{tabular}{ c | c c}
parameters      &   ULYSSES           &   CASSINI         \\
\hline
$(1/T)/{\rm Hz}$           &  $2\times 10^{-4}$   &   $10^{-4}$       \\
$S_0/{\rm Hz}^{-1}$    &  $ 2.5\times 10^{-24}$        &  $ 9\times 10^{-26}$ \\
\hline
$f_{12}/{\rm Hz}$        &     $2\times 10^{-4}$    &   $10^{-4}$    \\
$f_{23}/{\rm Hz}$        &     $5\times 10^{-2}$    &   $10^{-1}$    \\
\hline
$f_1/{\rm Hz}$           &   $2\times 10^{-4}$         &   $7.07\times 10^{-5}$         \\
$f_2/{\rm Hz}$           &   $8\times 10^{-4}$         &   $2.5\times 10^{-5}$         \\
$f_3/{\rm Hz}$  &   $5\times 10^{-2}$ &   $ 7.95\times 10^{-1}    $  \\
\hline
$\alpha_1$      &   -10                &        -2         \\
$\alpha_2$      &   -1/2                &        -1/2          \\
$\alpha_3$      &   10                &        2         \\
\end{tabular}
\end{table}

\begin{table}[c]
\caption{\label{tab;systems}
Main properties of the four fiducial sources used in the text. The assignment of the initial
frequency $f_b$, in the case of equal masses and coalescence at the 
end of the record, determines the final frequency $f_e$ (for a 40 days
run), the chirp mass $\Mc$ and the number of wave cycles ${\cal N}$ recorded at the 
instrument while the signal sweeps the frequency interval $(f_b,f_e)$.
}
\begin{tabular}{ c c c c}
$f_b / {\rm Hz}$ & $f_e / {\rm Hz}$ & $\Mc / \Ms$ & ${\cal N}$\\
\hline
$10^{-5}$          & $5.3\times 10^{-5}$  &  $3.6\times 10^7 $ & 52 \\
$5\times 10^{-5}$  & $6.9\times 10^{-4}$  &  $2.9\times 10^6 $ & 273 \\
$10^{-4}$          & $2.1\times 10^{-3}$  &  $9.2\times 10^5  $ & 550 \\
$5\times 10^{-4}$  & $2.7\times 10^{-2}$  &  $7.1\times 10^4   $ & 2761 \\
\end{tabular}
\end{table}

%
%
%

%
%

%
%
%
%

%
%

%
%

%
%

%
%

%
%

%
%

%
%

%
%

%
%
%
%

%
%

%
%
%
%

%
%

%
%


\begin{references}
\bibitem{PNapprox}
In order to extract the signal in a reliable and accurate way, the wave phase 
needs to be modelled including post-Newtonian correction terms \cite{will94,blanchet96}.
The phase evolution is currently known, in the finite mass case, up to the post$^{2.5}$-Newtonian
order \cite{BDIWW,BDI,WW,BIWW,B96}.
The post$^3$-Newtonian terms seem also to be under control now \cite{B98,BIJ98}.
\bibitem{note_class}
The inequalities in Eqs. (\ref{periodic}) -- (\ref{general})
are slightly pessimistic because they consider
the total frequency change from the beginning frequency, while one could
consider the frequency change from the middle frequency and obtain
a value about two times smaller, as we have done in \cite{pap1}.
\end{references}
\end{document}